\documentclass[11pt]{article}
\usepackage[left=2cm,top=2cm, bottom=2in, right=2cm]{geometry}
\date{}

\usepackage{enumitem}
\usepackage{amsfonts}
\usepackage{amsmath, amsthm, amssymb, dsfont}
\usepackage{graphicx}
\usepackage{hyperref, comment, cite}
\usepackage{psfrag}
\usepackage{bbm}
\usepackage{appendix}
\usepackage{csquotes}
\usepackage[table,dvipsnames]{xcolor}

\usepackage{tikz}
\usetikzlibrary{plotmarks}
\usetikzlibrary{quotes,arrows,decorations.pathmorphing,backgrounds,fit,petri,positioning,matrix,math}
\usepackage{pgfplots}
\usetikzlibrary{calc}%,fillbetween
\usetikzlibrary{shapes,arrows}
\usetikzlibrary{decorations.markings}
\usetikzlibrary{shapes.multipart} % multiline text in nod
\newcommand*\xor{\mathbin{\oplus}}
\tikzset{XOR/.style={draw,circle,append after command={
			[shorten >=\pgflinewidth, shorten <=\pgflinewidth,]
			(\tikzlastnode.north) edge (\tikzlastnode.south)
			(\tikzlastnode.east) edge (\tikzlastnode.west)
		}
	}
}
\tikzset{line/.style={draw, -latex',shorten <=1bp,shorten >=1bp}}

\usepackage{dsfont}
\usepackage{array,bm}
\newcolumntype{L}[1]{>{\raggedright\let\newline\\\arraybackslash\hspace{0pt}}m{#1}}
\newcolumntype{C}[1]{>{\centering\let\newline\\\arraybackslash\hspace{0pt}}m{#1}}
\newcolumntype{R}[1]{>{\raggedleft\let\newline\\\arraybackslash\hspace{0pt}}m{#1}}

\makeatletter
\newsavebox\myboxA
\newsavebox\myboxB
\newlength\mylenA
\newcommand*\xoverline[2][0.75]{%
	\sbox{\myboxA}{$\m@th#2$}%
	\setbox\myboxB\null% Phantom box
	\ht\myboxB=\ht\myboxA%
	\dp\myboxB=\dp\myboxA%
	\wd\myboxB=#1\wd\myboxA% Scale phantom
	\sbox\myboxB{$\m@th\overline{\copy\myboxB}$}%  Overlined phantom
	\setlength\mylenA{\the\wd\myboxA}%   calc width diff
	\addtolength\mylenA{-\the\wd\myboxB}%
	\ifdim\wd\myboxB<\wd\myboxA%
	\rlap{\hskip 0.5\mylenA\usebox\myboxB}{\usebox\myboxA}%
	\else
	\hskip -0.5\mylenA\rlap{\usebox\myboxA}{\hskip 0.5\mylenA\usebox\myboxB}%
	\fi}
\makeatother

\newcommand{\Enc}{\mathsf{Enc}}
\newcommand{\Dec}{\mathsf{Dec}}

\usepackage{accents}
\newcommand\thickbar[1]{\accentset{\rule{.8em}{0.8pt}}{#1}}

\makeatletter
\newcommand*\bigcdot{\mathpalette\bigcdot@{.5}}
\newcommand*\bigcdot@[2]{\mathbin{\vcenter{\hbox{\scalebox{#2}{$\m@th#1\bullet$}}}}}
\makeatother

\newtheorem{theorem}{Theorem}%[section]

\newtheorem{definition}{Definition}

\newtheorem{remark}{Remark}

\title{Secret Key Agreement with Physical Unclonable Functions: \\An Optimality Summary}

\author{Onur~G\"unl\"u and Rafael F. Schaefer% <-this % stops a space
	\thanks{O. G\"unl\"u and R. F. Schaefer were supported by the German Federal Ministry of Education and Research (BMBF) within the national initiative for ``Post Shannon Communication (NewCom)'' under the Grant 16KIS1004.}
	\thanks{The authors are with the Information Theory and Applications Chair, Technische Universit\"at Berlin, 10623 Berlin, Germany (e-mail: \{guenlue, rafael.schaefer\}@tu-berlin.de).}
}

\begin{document}
	\maketitle

\begin{abstract}
We address security and privacy problems for digital devices and biometrics from an information-theoretic optimality perspective, where a secret key is generated for authentication, identification, message encryption/decryption, or secure computations. A \textit{physical unclonable function (PUF)} is a promising solution for local security in digital devices and this review gives the most relevant summary for information theorists, coding theorists, and signal processing community members who are interested in optimal PUF constructions. Low-complexity signal processing methods such as transform coding that are developed to make the information-theoretic analysis tractable are discussed. The optimal trade-offs between the secret-key, privacy-leakage, and storage rates for multiple PUF measurements are given. Proposed optimal code constructions that jointly design the vector quantizer and error-correction code parameters are listed. These constructions include modern and algebraic codes such as polar codes and convolutional codes, both of which can achieve small block-error probabilities at short block lengths, corresponding to a small number of PUF circuits. Open  problems in the PUF literature from a signal processing, information theory, coding theory, and hardware complexity perspectives and their combinations are listed to stimulate further advancements in the research on local privacy and security.
\end{abstract}

%%%%%%%%%%%%%%%%%%%%%%%%%%%%%%%%%%%%%%%%%%
%\setcounter{section}{-1} %% Remove this when starting to work on the template.
\section{Motivations}
Fundamental advances in cryptography were made in secret during the 20th century. One exception was Claude E. Shannon's paper ``Communication Theory of Secrecy Systems'' \cite{shannonsecrecy}. Until 1967, the literature on security was not extensive, but a book \cite{histreview} with a historical review of cryptography changed this trend \cite{crypto}. Since then, the amount of sensitive data to be protected against attackers has increased significantly. Continuous improvements in security are needed and every improvement creates new possibilities for attacks \cite{pufintheory}.

Recent hardware-intrinsic security systems, biometric secrecy systems, 5th generation of cellular mobile communication networks (5G) and beyond as well as the emerging internet of things (IoT) networks, have several salient characteristics that differentiate them from existing architectures. These include the deployment of large numbers of possibly low-complexity terminals with light or no infrastructure, stringent constraints on latency, and primary applications of data gathering, inference, and control. These unique characteristics call for a rethinking of some of the fundamentals of data communications and storage. For instance, these characteristics make it very challenging to provide adequate secrecy and privacy primitives. In particular, traditional cryptographic protocols, which require key distribution or certificate management, might not be suitable to support the diverse applications supported by these technologies and might not be able to assure the privacy of personal information intrinsic in the data collected by such applications. Furthermore, low-complexity terminals may not have the processing power to implement such protocols, and even if they do, latency tolerances may not permit the processing time needed for cryptographic operations. Similarly, traditional methods of storing a secret key in a secure non-volatile memory (NVM) can be shown to be not secure due to possible invasive attacks to the hardware. Thus, secrecy and privacy for information systems are issues that need to be rethought in the context of recent networks, digital circuits, and database storage. 

Information-theoretic security is an emerging approach to provide secrecy and privacy for, e.g., wireless communication systems and networks by exploiting the unique characteristics of the wireless communication channel. Information-theoretic security methods such as physical layer security (PLS) use signal processing, advanced coding, and communication techniques to secure wireless communications at the physical layer. There are two key advantages of PLS. Firstly, it enables the use of resources available at the physical layer such as multiple measurements, channel training mechanisms, power, and rate control, which cannot be utilized by the upper layers of the protocol stack. Secondly, it is based on an information-theoretic foundation for secrecy and privacy that does not make assumptions on the computational capabilities of adversaries, unlike cryptographic primitives. By considering the security and privacy requirements of recent digital systems and the potential benefits from information-theoretic security and privacy methods, it can be seen that information-theoretic methods can complement or even replace conventional cryptographic protocols for wireless networks, databases, and user authentication and identification. Since information-theoretic methods do not generally require pre-shared secret keys, they might considerably simplify the key management in complicated networks. Thus, these methods might be able to fulfill the stringent hardware area constrains of digital devices and latency constraints in certain 5G/6G applications, or to save computations; hence, battery life for low-power devices such as IoT devices. Furthermore, information-theoretic methods offer ``built-in" secrecy and privacy, which are generally agnostic to the network infrastructure and provide better scalability as the size of a network or database increases.

\begin{figure}
	\centering
	\includegraphics[scale=.45]{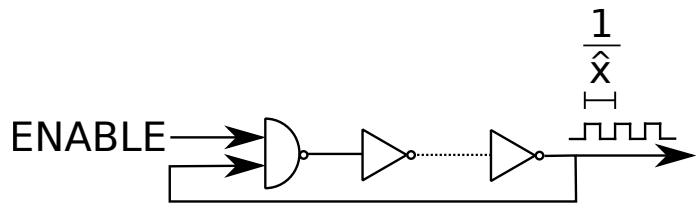}
	\caption{RO logic circuit.} \label{fig:ROlogiccircuit}
\end{figure} 

\begin{figure}
	\centering
	\includegraphics[width=0.32\textwidth, height=0.73\textheight, keepaspectratio=true]{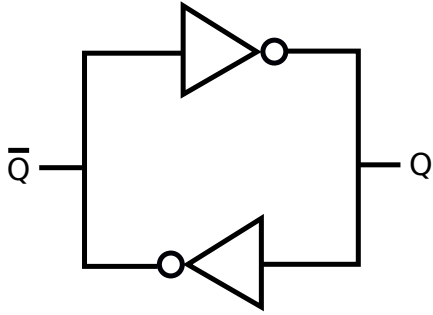}
	\caption{SRAM logic circuit.}
	\label{fig:SRAMlogiccircuit}
\end{figure}

A promising local solution to information-theoretic security and privacy problems is a \textit{physical unclonable function (PUF)} \cite{PUFFirst}. PUFs generate ``fingerprints'' for physical devices by using their intrinsic and unclonable properties. For instance, consider ring oscillators (ROs) with a logic circuit of multiple inverters serially connected with a feedback of the output of the last inverter into the input of the first inverter, as depicted in Figure~\ref{fig:ROlogiccircuit}. RO outputs are oscillation frequencies $1/\widehat{x}$, where $\widehat{x}$ is the oscillation period, that are unique and uncontrollable since the difference between different RO outputs is caused by submicron random manufacturing variations that cannot be controlled. One can use RO outputs as a source of randomness, called a \textit{PUF circuit}, to extract secret keys that are unique to the digital device that embodies these ROs. The complete method that puts out a unique secret key by using RO outputs is called an \textit{RO PUF}. Similarly, binary static random access memory (SRAM) outputs can be used as a source of randomness to implement SRAM PUFs in almost all digital devices because most digital devices have embedded SRAMs used for data storage. The logic circuit of an SRAM is depicted in Figure~\ref{fig:SRAMlogiccircuit} and the logically stable states of an SRAM cell are $(\overline{Q},Q)=(1,0)$ and $(0,1)$. During the power-up, the state is undefined if the manufacturer did not fix it. The undefined power-up state of an SRAM cell converges to one of the stable states due to random and uncontrollable mismatch of the inverter parameters, fixed when the SRAM cell is manufactured \cite{SoftHelper}. There is also random noise in the cell that affects the cell at every power-up. Since the physical mismatch of the cross-coupled inverters is a manufacturing variation, the power-up state of an SRAM cell is considered as a PUF response with one challenge, which is the address of the SRAM cell \cite{SoftHelper}.

PUFs resemble biometric features of human beings. In this review, we will list state-of-the-art methods that bridge the gap between the practical secrecy systems that use PUFs and the information-theoretic security limits by
\begin{itemize}
	\item Modeling real PUF outputs to solve security problems with valid assumptions;
	\item Analyzing methods that make information-theoretic analysis tractable, e.g., by transforming PUF symbols so that the transform-domain outputs are almost independent and identically distributed (i.i.d.), and that result in smaller hardware area than benchmark designs in the literature;
	\item Stating the information-theoretic limits for realistic PUF output models and providing optimal and practical (i.e., low-complexity and finite-length) code constructions that achieve these limits;
	\item Illustrating best-in-class nested codes for realistic PUF output models.
\end{itemize}
In short, we start with real PUF outputs to obtain mathematically-tractable models of their behaviour and then list optimal code constructions for these models. Since we discuss methods developed from the fundamentals of signal processing and information theory, any further improvements in this topic are likely to follow the listed steps in this review.

%%%%%%%%%%%%%%%%%%%%%%%%%%%%%%55
\subsection{Organization and Main Insights}
This paper is organized as follows. In Section~\ref{sec:pufbasics}, we define a PUF, list its existing and potential applications, and analyze the most promising PUF types. The PUF output models and design challenges faced when manufacturing reliable, low-complexity, and secure PUFs are listed in Section~\ref{sec:corrbiasnoisePUF}. The main security challenge in designing PUFs, i.e., output correlations, is tackled in Section~\ref{sec:transformcoding} mainly by using a transform coding method, which can provably protect PUFs against various machine learning attacks. The reliability and secrecy performance (e.g., the number of authenticated users) metrics used for PUF designs are defined and jointly optimized in Section~\ref{sec:quantandcodedesign}. PUF security and complexity performance evaluations for the defined transform coding method are given in Section~\ref{sec:comparisons}. Performance results for error-correction codes used in combination with previous code constructions that are used for key extraction with PUFs, are shown in Section~\ref{sec:correction} in order to illustrate that previous key extraction methods are strictly suboptimal. We next define the information theoretic metrics and the ultimate key-leakage-storage rate regions for the key agreement with PUFs problem, as well as comparing available code constructions for the key agreement problem in Section~\ref{sec:WZchapter}. Optimal code constructions for the key extraction with PUFs are implemented in Section~\ref{sec:codeproposal} by using nested polar codes, which are used in 5G networks in the control channel, to illustrate significant gains from using optimal code constructions. In Section~\ref{sec:DiscussionsandOpenProblems}, we provide a list of open PUF problems that might be interesting for information theorists, coding theorists, and signal processing researchers in addition to the PUF community.

%%%%%%%%%%%%%%%%%%%%%%%%%%%%%%%%%%%%%%%%%%%%%%%%%%
%%%%%%%%%%%%%%%%%%%%%%%%%%%%%%%%%%%%%%%%%%%%%%%%%%%
%%%%%%%%%%%%%%%%%%%%%%%%%%%%%%%%%%%%%%%%%%%%%%%%%%%
\section{PUF Basics}\label{sec:pufbasics}
We give a brief review of the literature on PUFs and discuss the problems with previous PUF designs that can be tackled by using signal processing and coding-theoretic methods. 

A PUF is a function that is embodied in a physical device and is unclonable. In the literature, there are alternative expansions of the term PUF such as ``physically unclonable function'', suggesting that it is a function that is only physically-unclonable. Such PUFs may provide a weaker security guarantee since they allow their functions to be digitally-cloned. For any practical application of a PUF, we need the property of unclonability both physically and digitally. We therefore consider a function as a PUF only if it is a physical function, which is embodied in a physical device, that is unclonable digitally and physically. 

Physical identifiers such as PUFs are heuristically defined to be complex challenge-response mappings that depend on the random variations in a physical object. Secret sequences are derived from this complex mapping, which can be used as a secret key. One important feature of PUFs is that the secret sequence generated is not required to be stored and it can be regenerated on demand. This property makes PUFs cheaper (no requirement for a memory for secret storage) and safer (the secret sequence is regenerated only on demand) alternatives to other secret generation and storage techniques such as storing the secret in an NVM \cite{PUFFirst}. 

There are an immense number of PUF types, which makes it practically impossible to give a single definition of PUFs that covers all types. We provide the following definition of PUFs that includes all PUF types of interest for this review.

\begin{definition}[\hspace{1sp}\cite{PUFFirst}]\label{def:PUFdefinition}
	A PUF is a challenge-response mapping embodied by a physical device such that it is fast and easy for the physical device to put out the PUF response and hard for an attacker, who does not have access the PUF circuits, to determine the PUF response to a randomly chosen challenge, even if he has access to a set of challenge-response pairs. 
\end{definition}  

The terms used in Definition~\ref{def:PUFdefinition}, i.e., fast, easy, and hard, are relative terms that should be quantified for each PUF application separately. There are physical functions, called physical one-way functions (POWFs), in the literature that are closely related to PUFs. Such functions are obtained by applying the cryptographic concept of ``one-way functions'', i.e., functions that are easy to evaluate but (on average) difficult to invert \cite{onewayfunction}, to physical systems. As the first example of POWFs, the speckle pattern obtained from coherent waves propagating through a disordered medium is a one-way function of both the physical randomness in the medium and the angle of the beam used to generate the optical waves \cite{PappuThesis}. 

Similar to POWFs, biometric identifiers such as the iris, retina, and fingerprints are closely related to PUFs. Most of the assumptions made for biometric identifiers are satisfied also by PUFs, so we can apply almost all of the results in the literature for biometric identifiers to PUFs. However, it is common practice to assume that PUFs can resist invasive (physical) attacks, which are considered to be the most powerful attacks used to obtain information about a secret in a system, unlike biometric identifiers that are constantly available for attacks. The reason for this assumption is that invasive attacks permanently destroy the fragile PUF outputs \cite{PUFFirst}. This assumption will be the basis for the PUF system models used throughout this review. We; therefore, assume that the attacker does not observe a sequence that is correlated with the PUF outputs, unlike biometric identifiers, since physical attacks applied to obtain such a sequence permanently change the PUF outputs.

% =====================================================================================================
\subsection{Applications of PUFs} \label{sec:application}
A PUF can be seen as a source of random sequences hidden from an attacker who does not have access to the PUF outputs. Therefore, any application that takes a secret sequence as input can theoretically use PUFs. We list some scenarios where PUFs fit well practically:

\begin{itemize}
	\item Security of information in wireless networks with an eavesdropper, i.e., a passive attacker, is a PLS problem. Consider Wyner's wiretap channel model introduced in \cite{WynerWTC}, where a transmitter sends a message through a broadcast channel so that a legitimate receiver can reliably reconstruct the message, while the message should be kept secret from an eavesdropper. This model is the most common PLS model, which is a channel coding problem unlike the secret key agreement problem we consider below that is a source coding problem. A randomized encoder helps the transmitter in keeping the message secret by confusing the eavesdropper. Therefore, one can use PUFs at the transmitter as the source of local randomness when a message should be sent securely through the wiretap channel. 
	\item Consider a 5G/6G mobile device that uses a set of SRAM outputs, which are available in mobile devices, as PUF circuits to extract secret keys so that the messages to be sent are encrypted with these secret keys before sending the data over the wireless channel. Thus, the receiver (e.g., a base station) that previously obtained the secret keys  (sent by mobile devices, e.g., via public key cryptography) can decrypt the data, while an eavesdropper who only overhears the data broadcast over the wireless channel cannot easily learn the message sent.   
	\item The controller area network (CAN) bus standard used in modern vehicles is illustrated in \cite{autonomouscars} to be susceptible to denial-of-service attacks, which shows that safety-critical inputs of the internal vehicle network such as brakes and throttle can be controlled by an attacker. One countermeasure is to encrypt the transmitted CAN frames by using block ciphers with secret keys generated from PUF outputs used as inputs.  
	\item IoT devices such as wearable or e-health devices may carry sensitive data and use a PUF to store secret keys so that only a mobile device with access to the secret keys can control the IoT devices. One common example of such applications is when PUFs are used to authenticate wireless body sensor network devices \cite{WBSNPUFs}.
	\item Cloud storage requires security to protect users' sensitive data. However, securing the cloud is expensive and the users do not necessarily trust the cloud service providers. A PUF in a universal serial bus (USB) token, i.e., Saturnus$^{\tiny{\textregistered}}$, has been trademarked to encrypt user data before uploading the data to the cloud, decrypted locally by reconstructing the same secret from the same PUF.
	\item System developers want to mutually authenticate a field programmable gate array (FPGA) chip and the intellectual property (IP) components in the chip, and IP developers want to protect the IP. In \cite{mutualauthenticatePUF}, a protocol is described to achieve these goals with a small hardware area that uses one symmetric cipher and one PUF. 
\end{itemize}   

Other applications of PUFs include providing non-repudiation (i.e., undeniable transmission or reception of data), proof of execution on a specific processor, and remote integrated circuit (IC) enabling. Every application of PUFs has different assumptions about the PUF properties, computational complexity, and the specific system models. Therefore, there are different constraints and system parameters for each application. We focus mainly on the application where a secret key is generated from a PUF for user, or device, authentication with privacy and secrecy guarantees, and low complexity. 

%%%%%%%%%%%%%%%%%%%%%%%%%%%%%%%%%%%%%%%%%%%%%%%
%%%%%%%%%%%%%%%%%%%%%%%%%%%%%%%%%%%%%%%%%%%%%%%%%%%
%%%%%%%%%%%%%%%%%%%%%%%%%%%%%%%%%%%%%%%%%%%%%%
\subsection{Main PUF Types}
We review four PUF types, i.e., silicon, arbiter, RO, and SRAM PUFs. We consider mainly the last two PUF types for algorithm and code designs due to their common use in practice and because signal processing techniques can tackle the problems arising in designing these PUFs. For a review of other PUF types that are mostly considered in the hardware design and computer science literatures, and various classifications of PUFs, see, e.g., \cite{DevadasPUFReview,TimPUFReview, pufintheory}. The four PUF types considered below can be shown to satisfy the assumption that invasive attacks permanently change PUF outputs, since digital circuit outputs used as the source of randomness in these PUF types change permanently under invasive attacks due to their dependence on nano-scale alterations in the hardware.

%%%%%%%%%%%%%%%%%%%%%%%%%%%%%%%%%%%%%%%%%%%%%%
\subsection{Silicon and Arbiter PUFs}
Common complementary metal-oxide-semiconductor (CMOS) manufacturing processes are used to build silicon PUFs, where the response of the PUF depends on the circuit delays which vary across integrated circuits (ICs) \cite{PUFFirst}. Due to high sensitivity of the circuit delays to environmental changes (e.g., ambient temperature and power supply voltage), arbiter PUFs are proposed in \cite{ExtractingSecretKeys}, for which an arbiter (i.e., a simple transparent data latch) is added to the silicon PUFs so that the delay comparison result is a single bit. The difference of the path delays is mapped to, e.g., the bit 0 if the first path is faster, and the bit 1 otherwise. The difference between the delays can be small, causing meta-stable outputs. Since the output of the mapper is generally preset to 0, the incoming signals must satisfy the setup time ($t_{\text{setup}}$) of the latch to switch the output to 1, resulting in a bias in the arbiter PUF outputs. Symmetrically implementable latches (e.g., set-reset latches) should be used to overcome this problem, which is difficult because FPGA routing does not allow the user to enforce symmetry in the hardware implementation. We discuss below that PUFs without symmetry requirements, e.g., RO PUFs, provide better results.

%%%%%%%%%%%%%%%%%%%%%%%%%%%%%%%%%
\subsection{Ring Oscillator PUFs} \label{subsec:ROPUFs}
The logic circuit of an odd number of inverters serially connected, where the output of the last inverter is given as input to the first inverter is an RO, as depicted in Figure~\ref{fig:ROlogiccircuit}. The first logic gate in Figure~\ref{fig:ROlogiccircuit} is a NAND gate, giving the same logic output as an inverter gate when the ENABLE signal is 1 (ON), to enable/disable the RO circuit. The manufacturing-dependent and uncontrollable component in an RO is the total propagation delay of an input signal to flow through the RO, determining the oscillation frequency $1/\hat{x}$ of an RO that is used as the source of randomness. A self-sustained oscillation is possible when the ring provides a 2$\pi$ phase shift and has unit voltage gain at the oscillation frequency $1/\hat{x}$. 

Consider an RO with $m\geq 3$ inverters. Each inverter should provide a phase shift of $\frac{\pi}{m}$ with an additional phase shift of $\pi$ due to the feedback. Therefore, the signal should flow through the RO twice to provide the necessary phase shift \cite{RingOscillators}. Suppose a propagation delay of $\tau_{d}$ for each inverter, so the oscillation frequency of an RO is $\displaystyle \frac{1}{\hat{x}} = \frac{1}{2 m \tau_d}$. We remark that since RO outputs are generally measured by using 32-bit counters, it is realistic to assume that a measured RO output  $\displaystyle \frac{1}{\hat{x}}$ is a realization of a continuous distribution that can be modeled by using the histogram of a family of RO outputs with the same circuit design, as assumed below.

\begin{figure}
	\centering
	\includegraphics[width=1.0\textwidth, height=0.73\textheight, keepaspectratio=true]{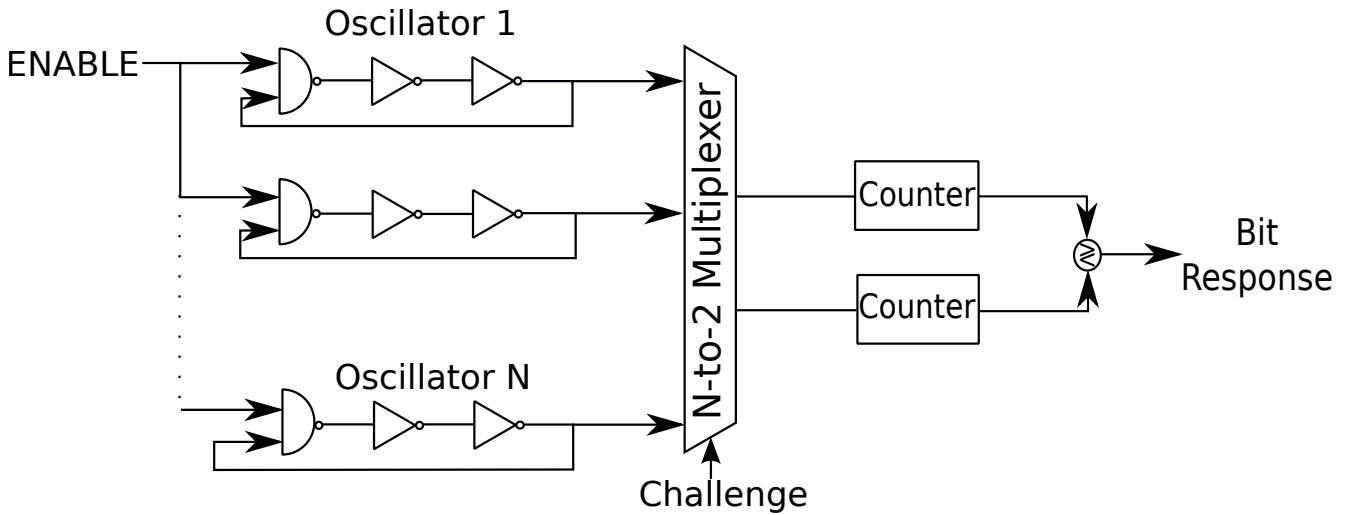}
	\caption{The first and most common RO PUF design \cite{ROFirst}.} \label{fig:ROPUFfirst}
\end{figure}

The propagation delay $\tau _{d}$ is affected by nonlinearities in the digital circuit. Furthermore, there are deterministic noise sources such as the cross-talk between adjacent signal traces and additional random noise sources such as thermal noise and flicker noise \cite{RingOscillators}. Such effects should be eliminated to have a reliable RO output. Rather than improving the standard RO designs, which would impose the condition that manufacturers should change their RO designs, the first proposal to fix the reliability problem was to make hard bit decisions by comparing RO pairs \cite{ROFirst}, as illustrated in Figure~\ref{fig:ROPUFfirst}.

In Figure~\ref{fig:ROPUFfirst}, the multiplexers are challenged by a bit sequence of length at most $\lceil\log _2 N \rceil$ so that an RO pair is selected among $N$ ROs. The counters put out the number of rising edges from each RO for a fixed time duration. A logic bit decision is made by comparing the counter values, which can be bijectively mapped to the oscillation frequencies. For instance, when the upper RO has a greater counter value, then the bit $0$ is generated; otherwise, the bit $1$. Given that ROs are identically laid out in the hardware, the differences in the oscillation frequencies are determined mainly by uncontrollable manufacturing variations. Furthermore, it is not necessary to have a symmetric layout when hard-macro hardware designs are used for different ROs, unlike arbiter PUFs. 

The key extraction method illustrated in Figure~\ref{fig:ROPUFfirst} gives an output of $N\choose 2$ bits, which are correlated due to overlapping RO comparisons. This causes a security threat and makes the RO PUF vulnerable to various attacks, including machine learning attacks. Thus, non-overlapping pairs of ROs are used in \cite{ROFirst} to extract each bit. However, there are systematic variations in the neighboring ROs due to the surrounding logic, which also should be eliminated to extract sequences with full entropy. Furthermore, ambient temperature and supply voltage variations are the most important effects that reduce the reliability of RO PUF outputs. A scheme called \textit{1-out-of-k masking} is proposed as a countermeasure to these effects, which compares the RO pairs that have the maximum oscillation frequency differences for a range of voltages and temperatures to extract bits \cite{ROFirst} . The bits extracted by such a comparison are more reliable than the bits extracted by using previous methods. The main disadvantages of this scheme are that it is inefficient due to unused RO pairs, and only a single bit is extracted from the (semi-) continuous RO outputs. We review transform-coding based RO PUF methods below that significantly improve on these methods without changing the standard RO hardware designs.

%%%%%%%%%%%%%%%%%%%%%%%%%%%%%%%%%%%%%%%%
\subsection{SRAM PUFs} 
There are multiple memory-based PUFs such as SRAM, Flip-flop, DRAM, and Butterfly PUFs. Their common feature is to posses a small number of challenge-response pairs with respect to their sizes. As the most promising memory-based PUF type that is already used in the industry, we consider SRAM PUFs that use the uncontrollable settling state of bi-stable circuits \cite{SRAM-PUF}. In the standard SRAM design, there are four transistors used to form the logic of two cross-coupled inverters, as depicted in Figure~\ref{fig:SRAMlogiccircuit}, and two other transistors to access the inverters. The power-up state, i.e., $(\overline{Q},Q)=(1,0)$ or $(0,1)$, of an SRAM cell provides one secret bit. Concatenating many such bits allows to generate a secret key from SRAM PUFs on demand. We provide an open problem about SRAM PUFs in Section~\ref{sec:DiscussionsandOpenProblems}.

%%%%%%%%%%%%%%%%%%%%%%%%%%%%%%%%%%%%%%%%%%%%%%%%%%
%%%%%%%%%%%%%%%%%%%%%%%%%%%%%%%%%%%%%%%%%%%%%%%%%%
%%%%%%%%%%%%%%%%%%%%%%%%%%%%%%%%%%%%%%%%%%%%%%%%%%
\section{Correlated, Biased, and Noisy PUF Outputs}\label{sec:corrbiasnoisePUF}
PUF circuit outputs are biased (nonuniform), correlated (dependent), and noisy (erroneous). We review a transform-coding algorithm that extracts an almost i.i.d. uniform bit sequence from each PUF, so a helper-data generation algorithm can correct the bit errors in the sequence generated from noisy PUF outputs. Using this transform-coding algorithm, we also obtain memoryless PUF measurement-channel models, so standard information-theoretic tools, which cannot be easily applied to correlated sequences, can be used. 

\begin{remark}
	The bias in the PUF circuit outputs is considered in the PUF literature to be a big threat against the security of the key generated from PUFs since the bias allows to apply, e.g., machine learning attacks. However, it is illustrated in \cite[Figure 6]{OnurProblem} that the output bias does not change the information-theoretic rate regions significantly, illustrating that there exist code constructions that do not require PUF outputs to be uniformly distributed.
\end{remark}

There are multiple \textit{key-generation}, i.e., \textit{generated-secret (GS)}, and \textit{key-binding}, i.e., \textit{chosen-secret (CS)}, methods to reconstruct secret keys from noisy PUF outputs, where the key is generated from the PUF outputs or bound to them, respectively. A code-offset fuzzy extractor (COFE) \cite{Dodis2008fuzzy} is an example of key-generation methods and the fuzzy-commitment scheme (FCS) \cite{FuzzyCommitment} is a key-binding method. Since a secret key should be stored in a secure database for both models, it might be practical to allow a trusted entity to choose the secret key that is bound to a PUF output. Thus, we first analyse a method that significantly improves reliability, privacy, secrecy, and hardware cost performance by using a transform-coding algorithm that is applied to PUF outputs in combination with the FCS. We remark that the information-theoretic analysis of the CS model follows directly from the analysis of the GS model \cite{IgnaTrans}, so one can use either model for comparisons.

Correlation in PUF outputs might leak information about the secret key, called \textit{secrecy leakage}, and about the PUF output, called \textit{privacy leakage} \cite{benimdissertation,IgnaTrans,ourMMMM}. Moreover, noise reduces the reliability of PUF outputs and error-correction codes are needed to satisfy the stringent reliability constraints. The transform-coding approach proposed in \cite{bizimpaper} in combination with a set of scalar quantizers has made its way into secret-key binding with continuous-output identifiers. This approach allows to significantly reduce the output correlation and to adjust the effective noise at the PUF output with reliability guarantees; see \cite{DPwithEfe} for a similar decorrelation approach applied to provide differential privacy to temporally-correlated eye movement measurements.  

%%%%%%%%%%%%%%%%%%%%%%%%%%%%%%%%%%%%%%%%%%%%%%%%%%%%%%%%%%%%%%%%%
%                   System Model
%%%%%%%%%%%%%%%%%%%%%%%%%%%%%%%%%%%%%%%%%%%%%%%%%%%%%%%%%%%%%%%%%%
\subsection{PUF Output Model}\label{sec:systemmodel} 
Consider a (semi-)continuous output physical function such as an RO output as a source that puts out a {real-valued} symbol $\hat{x}$. Systematic variations in RO outputs in a two-dimensional array are less than the systematic variations in one-dimensional ROs, since in a two-dimensional array the maximum distance between RO hardware logic circuits is less, which decreases the variations in the RO outputs caused by surrounding hardware logic circuits \cite{maiti2011improved}. Thus, consider a two-dimensional RO array of size $\displaystyle l\!= r\!\times\! c$ and represent the array as a vector random variable $\widehat{X}^l$. Suppose there is a single two-dimensional RO array in each device with the same circuit design and the RO array emits an output $\displaystyle \widehat{X}^l$ according to a probability density $f_{\widehat{X}^l}$. Each RO output is disturbed by mutually-independent additive Gaussian noise and the vector noise is denoted as $\widehat{Z}^l$. Define the noisy RO outputs as $\widehat{Y}^l\! =\! \widehat{X}^l \!+\! \widehat{Z}^l$. Observe~that $\widehat{X}^l$ and $\widehat{Y}^l$ are correlated. A secret key can thus be agreed by using these outputs \cite{AhlswedeCsiz,Maurer}. 
 
	\begin{remark}
		PUF outputs are noisy, as discussed above in this section. However, the first PUF outputs are used by, e.g., a manufacturer to generate or embed a secret key, which is called the \emph{enrollment} procedure. Since a manufacturer can measure multiple noisy outputs of the same RO to estimate the noiseless RO output, we can consider that the PUF outputs measured during enrollment are noiseless. However, during the reconstruction step, e.g., an IoT device observes a noisy RO output, which can be the case because the IoT device cannot measure the RO outputs multiple times due to delay and complexity constraints. Therefore, we consider a key-agreement model where the first measurement sequence (during enrollment) is noiseless and the second measurement sequence (during reconstruction) is noisy; see also Figure~\ref{fig:problemsetup} below. Extensions to key agreement models with two noisy sequences, where the noise components can be correlated, are discussed in \cite{ourMMMM,ourISITPoor,ourBCITW}.
	\end{remark}

One needs to extract random sequences with i.i.d. symbols from $\widehat{X}^l$ and $\widehat{Y}^l$ to employ available information-theoretic results in \cite{IgnaFuzzy} for secret-key binding with identifiers by using the FCS. An algorithm is proposed in \cite{bizimpaper} that extracts almost i.i.d. binary and uniformly distributed random vectors $X^n$ and $Y^n$ from $\widehat{X}^l$ and $\widehat{Y}^l$, respectively. For such $\displaystyle X^n$ and $\displaystyle Y^n$, we can define a binary error vector as $E^n\! =\! X^n\! \xor\! Y^n$, where $\xor$ is the modulo-2 sum. The random sequence $\displaystyle E^n$ corresponds to a sequence of i.i.d. Bernoulli random variables with parameter $p$, {i.e.,} $E^n\sim\text{Bern}^n(p)$. The channel $P_{Y|X}$ is thus a binary symmetric channel (BSC) with crossover probability $p$, i.e., BSC($p$). We discuss a transform-coding method below, which further provides reliability guarantees for each bit generated.

\begin{figure}
	\centering
	\resizebox{0.55\linewidth}{!}{
		\begin{tikzpicture}
		\node (a) at (0,-1.5) [XOR,scale=1.0] {};
		\node (b) at (6,-1.5) [XOR,scale=1.0] {};
		% Enrollment Box
		\node (f) at (0,-0.7) [draw,rounded corners = 6pt, minimum width=2.8cm,minimum height=3cm, align=left] {};
		% Reconstruction Box
		\node (g) at (6,-0.7) [draw,rounded corners = 6pt, minimum width=2.75cm,minimum height=3cm, align=left] {};
		% Enc Box
		\node (d) at (0,-0.1) [draw,rounded corners = 6pt, minimum width=2.6cm,minimum height=0.7cm, align=left] {$
			C^n = \Enc\left(S\right)$};
		% Channel Box
		\node (c) at (3,-2.7) [draw,rounded corners = 5pt, minimum width=1.3cm,minimum height=0.65cm, align=left] {$P_{Y|X}$};
		% Dec Box
		\node (e) at (6,-0.1) [draw,rounded corners = 6pt, minimum width=2.6cm,minimum height=0.7cm, align=left] {$\hat{S} = \Dec\left(R^n\right)$};
		% Enc to Dec Arrow
		\draw[decoration={markings,mark=at position 1 with {\arrow[scale=1.5]{latex}}},
		postaction={decorate}, thick, shorten >=0.6pt] (a.east) -- (b.west) node [midway, above] {$W$};
		% X^N goes to Enc Box
		\node (a1) [below of = a, node distance = 1.2cm] {$X^n$};
		\node (b1) [below of = b, node distance = 1.2cm] {$Y^n$};
		% X^N  to Enc arrow
		\draw[decoration={markings,mark=at position 1 with {\arrow[scale=1.5]{latex}}},
		postaction={decorate}, thick, shorten >=1.4pt] (a1.north) -- (a.south);
		% X^N  to Channel box arrow
		\draw[decoration={markings,mark=at position 1 with {\arrow[scale=1.5]{latex}}},
		postaction={decorate}, thick, shorten >=1.4pt] (a1.east) -- (c.west);
		% Channel box to Y^N arrow
		\draw[decoration={markings,mark=at position 1 with {\arrow[scale=1.5]{latex}}},
		postaction={decorate}, thick, shorten >=1.4pt] (c.east) -- (b1.west);
		% Y^N  to Enc arrow
		\draw[decoration={markings,mark=at position 1 with {\arrow[scale=1.5]{latex}}},
		postaction={decorate}, thick, shorten >=1.4pt] (b1.north) -- (b.south);
		% S exchanges with Enc
		\node (a2) [above of = d, node distance = 1.5cm] {$S$};
		% Enrollment
		\node (f2) [below of = f, node distance = 2.5cm] {Enrollment};
		% Reconstruction
		\node (g2) [below of = g, node distance = 2.5cm] {Reconstruction};
		% \hat{S} exchanges with Dec
		\node (b2) [above of = e, node distance = 1.5cm] {$\hat{S}$};
		% Dec to \hat{S} arrow
		\draw[decoration={markings,mark=at position 1 with {\arrow[scale=1.5]{latex}}},
		postaction={decorate}, thick, shorten >=1.4pt] (e.north) -- (b2.south);
		% Enc to S arrows
		% \draw[decoration={markings,mark=at position 1 with {\arrow[scale=1.5]{latex}}},
		%     postaction={decorate}, thick, densely dotted, shorten >=1.4pt] ([xshift=-10pt]a2.south) -- ([xshift=-10pt]a.north);
		\draw[decoration={markings,mark=at position 1 with {\arrow[scale=1.5]{latex}}},
		postaction={decorate}, thick, shorten >=1.4pt] (a2.south) -- (d.north); 
		% XOR2 to Dec arrow
		\draw[decoration={markings,mark=at position 1 with {\arrow[scale=1.5]{latex}}},
		postaction={decorate}, thick, shorten >=1.4pt] (b.north) -- (e.south) node [midway, right] {$R^n$};
		% XOR1 to Enc arrow
		\draw[decoration={markings,mark=at position 1 with {\arrow[scale=1.5]{latex}}},
		postaction={decorate}, thick, shorten >=1.4pt]  (d.south) -- (a.north) node [midway, left] {$C^n$};;
		\end{tikzpicture}
	}
	\caption{The fuzzy commitment scheme (FCS).}\label{fig:fuzzycommitment}
\end{figure}
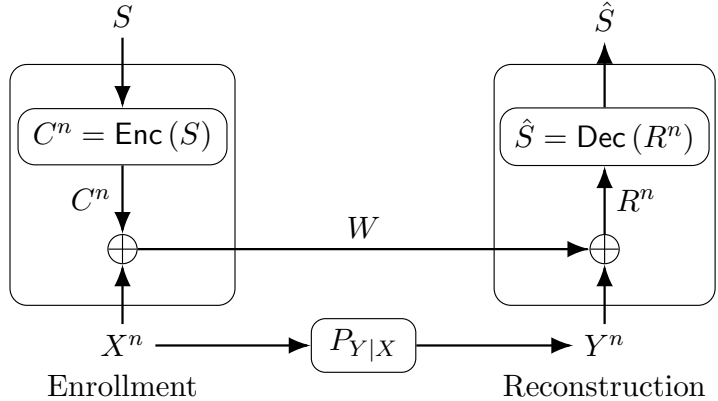

The FCS can reconstruct a secret key by using correlated random variables without leaking any information about the secret key \cite{FuzzyCommitment}. The FCS is depicted in Figure~\ref{fig:fuzzycommitment}, where an encoder $\Enc(\cdot)$  maps a secret key $S\in\mathcal{S}$, which is uniformly distributed in the set $\{1,2,\ldots,|\mathcal{S}|\}$, into a binary codeword $\displaystyle C^n$ that is added modulo-2 to the binary PUF-output sequence $\displaystyle X^n$ during enrollment. The resulting sequence is called helper data $\displaystyle W$, sent through a public and noiseless communication link to a database. The modulo-2 sum of the helper data $W$ and $Y^n$ gives the result $R^n=W\xor Y^n=C^n\!\xor E^n$, which can be later mapped to an estimate $\displaystyle \hat{S}$ of the secret key $S$ by the decoder $\displaystyle \Dec(\cdot)$ during reconstruction.

We next give information-theoretic rate regions for the FCS; see \cite{CoverandThomas} for information-theoretic notation and basics.  
%%%%%%%%%%%%%%%%%%%%%%%%%%%%%%%%%%%%%%%%%%%%%%%%%%%%%%%%%%%%%%%%%%%%%%%%%%%
%%%%%%%%%%%%%%%%%%%%%%%%%%%%%%%%%%%%%%%%%%%%%%%%%%%%%%%%%%%%%%%%%%%%%%%%%%%%%%%%%%

\begin{definition}\label{def:ratepair}
	A secret-key vs. privacy-leakage rate pair $\displaystyle \left(R_\text{s}\!\;\text{,}\;\!R_\ell\right)$ is achievable by the FCS with perfect secrecy, {i.e.,} zero secrecy leakage, if, given any $\delta\!>\!0$, there is some $n\!\geq\!1$, and an encoder and decoder for which $\displaystyle R_\text{s}=\frac{\log|\mathcal{S}|}{n}$ and
	\begin{alignat}{2} 
	&P_B=\Pr[S\ne\hat{S}] \leq \delta && (\text{reliability})\\
	&H(S)\geq n(R_\text{s}-\delta)&&(\text{key uniformity})\\
	&I\left(S;W\right)\!=\!0 && (\text{perfect secrecy})\label{eq:secrecyconst}\\
	&I\left(X^n;W\right) \leq n(R_\ell+\delta) \quad\quad\quad&&(\text{privacy})\label{eq:privacyconstFCS}
	\end{alignat}
	where (\ref{eq:secrecyconst}) suggests that $S$ and $W$ are independent and (\ref{eq:privacyconstFCS}) suggests that the rate of dependency between $X^n$ and $W$ is bounded. The achievable secret-key vs. privacy-leakage rate, or key-leakage, region $\mathcal{R}_{\text{FCS}}$ for the FCS is the union of all achievable pairs.
\end{definition}

\begin{theorem}[\hspace{1sp}\cite{IgnaFuzzy}]
	The key-leakage region $\mathcal{R}_{\text{FCS}}$ for the FCS with a channel $P_{Y|X}$ that is a BSC$(p)$, uniformly distributed $X$ and $Y$, and zero secrecy leakage is 
	\begin{align}
	\mathcal{R}_{\text{FCS}} = \{&\left(R_\text{s},R_\ell\right)\colon\;\;\; 0\leq R_\text{s}\leq 1-H_b(p),\qquad R_\ell\geq 1-R_\text{s}\}\label{eq:ls0}
	\end{align}
	where $H_b(p)=-p\log p - (1-p)\log(1-p)$ is the binary entropy function.
\end{theorem}

The region $\mathcal{R}_{\text{FCS}}$ suggests that any (secret-key, privacy-leakage) rate pair that sums to 1~bit/source-bit is achievable with the constraint that the secret-key rate is at most the channel capacity of the BSC($p$), i.e., $\underset{p_X}{\max}\;I(X;Y)=1-H_b(p)$. Furthermore, smaller secret-key rates and greater privacy-leakage rates than these rates are also achievable. 

The FCS is a particular realization of the CS model. The region $\mathcal{R}$ of all achievable (secret-key, privacy-leakage) rate pairs for the CS model with a negligible secrecy-leakage rate, where a generic encoder is used to confidentially transmit an embedded secret key to a decoder that observes $Y^n$ and the helper data $W$, is given in \cite{IgnaTrans} as
\begin{align}
\mathcal{R}\! =\! &\bigcup_{P_{U|X}}\!\Bigg\{\left(R_\text{s},R_\ell\right)\!\colon\!\quad 0\leq R_\text{s}\leq I(U;Y),\qquad R_\ell\geq I(U;X)-I(U;Y)\Bigg\}\label{eq:chosensecret}
\end{align}
where $U-X-Y$ forms a Markov chain and the alphabet $\mathcal{U}$ of the auxiliary random variable $U$ can be limited to have the size $\displaystyle |\mathcal{U}|\!\leq\!|\mathcal{X}|+1$. The auxiliary random variable $U$ represents a distorted version of $X$ through a channel $P_{U|X}$. The FCS is optimal, {i.e.,}~it~achieves a boundary point of $\mathcal{R}$, for a BSC $P_{Y|X}$ with crossover probability $p$ only at the point $\displaystyle (R_{\text{s}}^*,R_\ell^*)\!=\!(1\!-\!H_b(p),H_b(p))$ \cite{IgnaFuzzy}. This point corresponds to the highest achievable secret-key rate; see Figure~\ref{fig:ratecomparison} below. We remark that the region $\mathcal{R}$ gives an outer bound for the perfect-secrecy case considered to obtain $\mathcal{R}_{\text{FCS}}$.

%%%%%%%%%%%%%%%%%%%%%%%%%%%%%%%%%%%%%%%%%%%%%%%%%%%%%%%%%%%%%%%%%%%%%%%%%%%%
%%                          Common Post Processing Steps
%%%%%%%%%%%%%%%%%%%%%%%%%%%%%%%%%%%%%%%%%%%%%%%%%%%%%%%%%%%%%%%%%%%%%%%%%%%
\section{Transform Coding  Steps}\label{sec:transformcoding}

\begin{figure}
	\centering
	\includegraphics[width=0.8\textwidth, height=0.8\textheight, keepaspectratio]{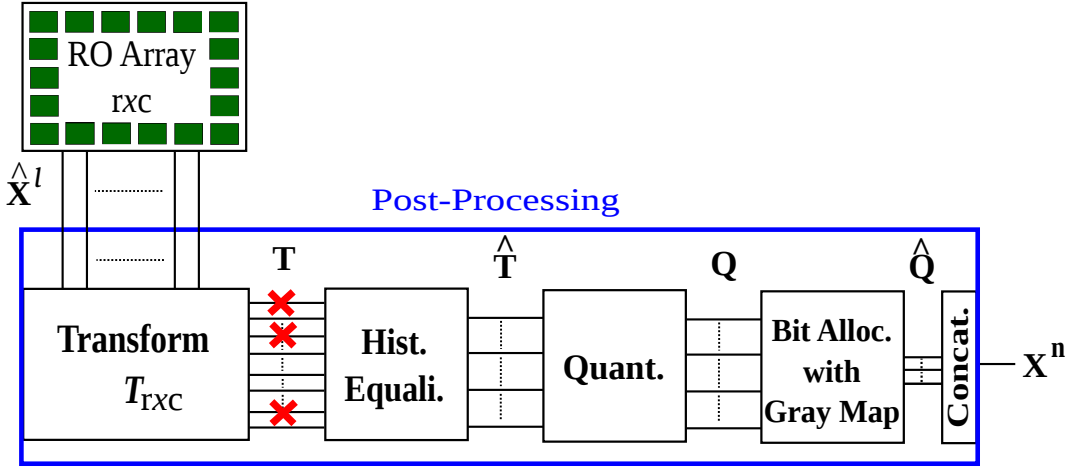}
	\caption{{Transform-coding steps \cite{bizimpaper}.}} 
	\label{fig:postprocessing}
\end{figure}

The main aim of transform coding is to reduce correlations between outputs of the ROs in the same two-dimensional array by using, e.g., a linear transformation. We discuss a transform-coding algorithm proposed in \cite{bizimMDPI} as an extension of \cite{bizimpaper} to provide reliability guarantees to each generated bit. Joint optimization of the quantizer and error-correction code parameters to maximize the security and reliability performance, and a simple method to decrease hardware storage are its main steps. The output of these post-processing steps is a bit sequence $X^n$ (or its noisy version $Y^n$) used in the FCS. One applies the same post-processing steps for the enrollment and reconstruction. The~difference is that during enrollment the design parameters are chosen as a function of the PUF-circuit output statistics by the device manufacturer. It thus suffices to discuss only the enrollment steps. Figure~\ref{fig:postprocessing} shows the post-processing steps that include transformation, histogram equalization, quantization, bit allocation, and bit-sequence concatenation.

RO outputs $\widehat{X}^l$ are correlated due to, {e.g.,} the surrounding logic in the hardware. A~transform $\emph{T}_{r\!\times\!c}(\cdot)$ of size $\displaystyle r\!\times\! c$ is applied to an array of RO outputs to reduce correlations. Decorrelation~performance of a transform depends on the source statistics. We model each {real-valued} output $T$ in the transform domain, called \emph{transform coefficient}, obtained from an RO-output dataset in \cite{ROLarge} by using the corrected Akaike's information criterion (AICc) \cite{CorrAIC} and the Bayesian information criterion (BIC) \cite{BIC2}. These criteria suggest that a Gaussian distribution can be fitted to each transform coefficient $T$ for the DCT, discrete Walsh-Hadamard transform (DWHT), discrete Haar transform (DHT), and Karhunen-Lo\`{e}ve transform (KLT), which are common orthogonal transforms considered in the literature for image processing, digital watermarking, etc. Use maximum-likelihood estimation to derive unbiased estimates for the parameters of Gaussian distributions. 

%%%%%%%%%%%%%%%%%%%%%%%%%%%%%%%%%%%%%
%%%%%%%%%%%%%%%%%%%%%%%%%%%%%%%%%%%%%%%%%

The histogram equalization step in Figure~\ref{fig:postprocessing} converts the probability density of the $i$-th coefficient $T_i$ into a standard normal distribution such that $ \widehat{T}_i = \frac{T_i - \mu_i}{\sigma_i}$, where $\displaystyle \mu_i$ is the mean and $\displaystyle \sigma_i$ is the standard deviation of the $i$-th transform coefficient for all $\displaystyle i\!=\!1,2,\ldots,l$. Quantization steps for all transform coefficients are thus the same. Without histogram equalization, we need a different quantizer for each transform coefficient. Therefore, the histogram equalization step reduces the storage for the quantization steps. Transformed and equalized coefficients $\displaystyle \widehat{T}_i$ are independent if the transform $\emph{T}_{r\!\times\!c}(\cdot)$ decorrelates the RO outputs perfectly and the transform coefficients $\displaystyle T_i$ are jointly Gaussian. One can thus use a scalar quantizer for all coefficients without a performance loss for this case. Scalar quantizers and bit extraction methods are given below to satisfy the security and reliability requirements of the FCS with the independence assumption, which can be combined with a correlation-thresholding approach in practice.

%%%%%%%%%%%%%%%%%%%%%%%%%%%%%%%%%%%%%%%%%%%%%%%%%%%%%%%
\section{Joint Quantizer and Error-Correction Code Design}\label{sec:quantandcodedesign}
The aim of the post-processing steps in Figure~\ref{fig:postprocessing} is to extract a uniformly-random bit sequence $\displaystyle X^n$. Use a quantizer $\displaystyle \Delta(\cdot)$ with quantization-interval values $\displaystyle k=1,2,\cdots,2^{K_i}$, where $\displaystyle K_i$ is the number of bits we extract from the $i$-th coefficient $\widehat{T}_i$ for $i\!=\!1,2,\ldots,l$. Let
\begin{align}
\Delta(\hat{t}_i) = k\quad \text{if}\quad  b_{k-1}\!<\!\hat{t}_i\!\leq\!b_k \label{eq:quantizer}
\end{align}
and choose $\displaystyle b_k = \Phi^{-1}\left(\frac{k}{2^{K_i}}\right)$, where $\displaystyle \Phi^{-1}(\cdot)$ is the quantile function of the standard normal distribution. The output $k$ is assigned to a bit sequence of length $K_i$. The most likely error event when we quantize $\widehat{T}_i$ is a jump to a neighboring quantization step due to zero-mean noise model assumed. Thus, apply a Gray mapping when assigning bit sequences of length $K_i$ to the integers $k=1,2,\ldots,2^{K_i}$, so neighboring bit sequences change only in one bit.

We next discuss a reliability metric proposed for a joint quantizer and code design by fixing the maximum number of erroneous transform coefficients and considering an error-correction code that can correct all error patterns with up to a fixed number of errors.

%%%%%%%%%%%%%%%%%%%%%%%%%%%%%%%%%%%%%%%%%%%%
\subsection{Quantizer Design with Fixed Maximum Number of Errors}
We discuss a conservative approach that assumes that either all bits extracted from a transform coefficient are correct or they all flip. The \textit{correctness probability} $P_c$ of a transform coefficient is defined to be the probability that all bits associated with this coefficient are correct. This metric is used to determine the number of bits extracted from each coefficient such that there is a channel encoder and a bounded minimum distance decoder (BMDD) that satisfy the block-error probability constraint $P_B\leq10^{-9}$, which is a common block-error probability considered for PUFs that consist of CMOS circuits \cite{ROFirst}. This approach results in reliability guarantees for the random-output RO arrays.

Let $Q(\cdot)$ be the Q-function, $\displaystyle \sigma^2_{\hat{n}}$ the noise variance, and $\displaystyle f_{\widehat{T}}$ the probability density of the standard Gaussian distribution. For a $K$-bit quantizer and the quantization boundaries $b_k$ as in (\ref{eq:quantizer}) for an equalized Gaussian transform coefficient $\widehat{T}$, the correctness probability is
\begin{align}
P_c&(K) \!=\!\sum_{k=0}^{2^K-1}\!\int\displaylimits_{b_{k}}^{b_{k+1}}\Bigg[\!Q\Big(\frac{b_{k}\!-\!\hat{t}}{\sigma_{\hat{n}}}\Big)\!-\!Q\Big(\frac{b_{k+1}\!-\!\hat{t}}{\sigma_{\hat{n}}}\Big)\!\Bigg]f_{\widehat{T}}(\hat{t})d{\hat{t}}\label{eq:correctness}
\end{align}
which calculates the probability that the additive noise will not change the quantization interval assigned to the transform coefficient, i.e., all bits associated with the transform coefficient stay the same after adding noise.

Suppose the channel decoder can correct all errors in up to $\displaystyle C_{\text{max}}$ transform coefficients. Suppose~further that coefficient errors occur independently, i.e., noise on different transform coefficients are mutually independent, and that the correctness probability $\displaystyle P_{c,i}(K)$ of the $i$-th coefficient $\widehat{T}_i$ for $i\!=\!1,2,\ldots,l$ is at least $\displaystyle \thickbar{P}_c(C_{\text{max}})$. A sufficient condition for satisfying the block-error probability constraint $P_B\!\leq\!10^{-9}$ is that $\displaystyle \thickbar{P}_c(C_{\text{max}})$ satisfies the inequality
\begin{align}
\sum_{c=C_{\text{max}}+1}^{l}{l \choose c}{(1\!-\!\thickbar{P}_c(C_{\text{max}}))}^{c}{\thickbar{P}_c(C_{\text{max}})}^{l-c}\!\leq\! 10^{-9}\label{eq:threshold}.
\end{align} 
 
Determine the number $K_i$ of bits extracted from the $i$-th transform coefficient as the maximum value $K$ such that $\displaystyle P_{c,i}(K)\geq \thickbar{P}_{c}(C_{\text{max}})$. The first coefficient, i.e., DC coefficient, $\widehat{T}_1$ is not used since its value is a scaled version of the mean of the RO outputs in the same array, which is generally known by an attacker. Ambient-temperature and supply-voltage variations have a highly-linear effect on the RO outputs, so the DC coefficient is the most affected coefficient, which is another reason not to use the DC coefficient \cite{bizimtemperature}. Therefore, choose $K_1\!=\!0$ so that the total number $\displaystyle n(C_{\text{max}})$ of extracted bits is
\begin{align}
n(C_{\text{max}})\!=\!\sum_{i=2}^l K_i \label{eq:totalbits}.
\end{align}
In the worst case, the coefficients in error are the coefficients from which the largest number of bits are extracted. Sort the numbers $K_i$ of bits extracted from all coefficients in descending order such that $\displaystyle K^{\prime}_{i}\!\geq\!K^{\prime}_{i+1}$ for all $i\!=\!1,2,\ldots,l-2$. The channel decoder thus must be able to correct up to
\begin{align}
e( C_{\text{max}}) = \sum_{i=1}^{C_{\text{max}}}  K^{\prime}_{i}\label{eq:minimume}
\end{align}
bit errors, which can be satisfied by using a block code with minimum distance $\displaystyle d_{\text{min}}\!\geq\!2e(C_{\text{max}})\!+\!1$ since a BMDD can correct all errors with up to $e=\lfloor\frac{d_{\text{min}-1}}{2}\rfloor$ errors \cite{ECC}. 

Suppose a key bound to physical identifiers in a device is used in the advanced encryption standard (AES) with a uniformly-distributed secret key of length 128 bits. The block code used in the FCS should thus have a code length of at most $\displaystyle n(C_{\text{max}})$ bits, code dimension of at least $128$ bits, and minimum distance of $\displaystyle d_{\text{min}} \geq 2e(C_{\text{max}})+1$ for a fixed $\displaystyle C_{\text{max}}$. The code rate should be as high as possible to operate close to the optimal (secret-key, privacy-leakage) rate point of the FCS. This optimization problem is hard to solve. One can illustrate by an exhaustive search over a set of $\displaystyle C_{\text{max}}$ values and over a selection of algebraic codes that there is a channel code that satisfies these constraints with a reliability guarantee for each extracted bit. Restricting the search to codes that admit low-complexity encoders and decoders is desired for, e.g., IoT applications, for which complexity is the bottleneck.

Conditions listed above are conservative. For a given transform coefficient, the correctness probability can be significantly greater than the correctness threshold $\displaystyle \thickbar{P}_c(C_{\text{max}})$. Secondly, due to Gray mapping, it is more likely that less than $K_i$ bits are in error when the $i$-th coefficient is erroneous. It is also unlikely that the bit errors always occur in the transform coefficients from which the largest numbers of bits are extracted. Thus, even if a channel code cannot correct all error patterns with up to $e( C_{\text{max}})$ errors, it can still be the case that the block-error probability constraint is satisfied. We next illustrate such a case.
%%%%%%%%%%%%%%%%%%%%%%%%%%%%%%%%%%%%%%%%%%%%%%%%%%%%%%%%%%%%%%%%%
%%                   Choice of the Parameters
%%%%%%%%%%%%%%%%%%%%%%%%%%%%%%%%%%%%%%%%%%%%%%%%%%%%%%%%%%%%%%%%%
\section{PUF Performance Evaluations}\label{sec:comparisons}
Suppose RO outputs $\widehat{X}^l$ are represented as a vector random variable with an autocovariance matrix $\mathbf{C_{\widehat{X}\widehat{X}}}$. Consider RO arrays of sizes $8$ $\!\times \!$ $8$ and 16 $\!\times \!$ 16. Autocovariance matrix of such RO array outputs and noise parameters can be estimated from the RO output dataset in \cite{ROLarge}. Using the dataset, we compare the performance of the DCT, DWHT, DHT, and KLT in terms of their decorrelation efficiency, complexity, uniqueness, and~security. 

%%%%%%%%%%%%%%%%%%%%%%%%%%%%%%%%%%%%%%%%
% Decorrelation Performance
%%%%%%%%%%%%%%%%%%%%%%%%%%%%%%%%%%%%%%%%%

\subsection{Decorrelation Performance}
One should eliminate correlations between the RO outputs and make them independent to extract uniform bit sequences by quantizing each transform coefficient separately. Use the \textit{decorrelation} \textit{efficiency} $\displaystyle \eta_c$ \cite{decorrelation} as a decorrelation performance metric. Consider the autocovariance matrix $\mathbf{C_{TT}}$ of the transform coefficients, so $\displaystyle \eta_c$ of a transform is 
\begin{align}
\eta_c = 1-\frac{\sum\limits_{a=1}^{l}\sum\limits_{b=1}^{l}|\mathbf{C_{TT}}(a,b)|\mathds{1}\{a\!\ne\!b\}}{\sum\limits_{a=1}^{l}\sum\limits_{b=1}^{l}|\mathbf{C_{\widehat{X}\widehat{X}}}(a,b)|\mathds{1}\{a\!\ne\!b\}}
\end{align}
where the indicator function $\displaystyle \mathds{1}\{a\!\ne\!b\}$ takes on the value 1 if $\displaystyle a\!\ne\!b$ and 0 otherwise. The decorrelation efficiency of the KLT is 1, which is optimal \cite{decorrelation}. We list the average decorrelation efficiency results of other transforms in Table~\ref{tab:decorrelationeff}. All transforms have similar and good decorrelation efficiency performance for the RO outputs in the dataset \cite{ROLarge}. The DCT and DHT have the highest efficiency for $\displaystyle 8\!\times \!8$ RO arrays, whereas for $\displaystyle 16 \!\times \! 16$ RO arrays, the best transform is the DWHT. Table~\ref{tab:decorrelationeff} indicates that increasing the array size improves $\displaystyle \eta_c$.

\begin{table}[t]
	\caption{The average RO output decorrelation-efficiency results.}
	\centering
	\begin{tabular}{ c|C{1.7cm}|C{1.7cm}|C{1.7cm}| }
		\cline{2-4}
		& DCT & DWHT & DHT\\
		\cline{1-4}
		\multicolumn{1}{|c|}{$\displaystyle \eta_c$ for $8\times 8$} &  0.9978 &  0.9977&  0.9978\\
		\cline{1-4}
		\multicolumn{1}{|c|}{$\displaystyle \eta_c$ for $16\times 16$} & 0.9987 &0.9988 &  0.9986\\
		\cline{1-4}
	\end{tabular}\label{tab:decorrelationeff}
\end{table}

%%%%%%%%%%%%%%%%%%%%%%%%%%%%%%%%%%%%%%%%%%%%%%%%%%%%
%           Complexity
%%%%%%%%%%%%%%%%%%%%%%%%%%%%%%%%%%%%%%%%%%%%%%%%%%%%
\subsection{Transform Complexity}
We measure the complexity of a transform in terms of the number of operations required to compute the transform. We are interested in a computational-complexity comparison for RO arrays of sizes $ r\!=\!c\!=\!8$ and $r\!=\!c\!=\!16$, which are powers of 2, so that fast algorithms are available for the DCT, DWHT, and DHT. The computational complexity of the KLT for $r\!=\!c\!=\!n$ is $\displaystyle O(n^3)$, while it is $\displaystyle O(n^2\log_2 n)$ for the DCT and DWHT, and $\displaystyle O(n^2)$ for the DHT \cite{mySPbook}. There are efficient implementations of the DWHT without multiplications \cite{bizimMDPI}, which can be applied also to the transforms proposed in \cite{bizimICASSP2020}. The DWHT is thus a good candidate for RO PUF designs for IoT applications. For instance, a hardware implementation of two-dimensional (2D) DWHT in a Xilinx ZC706 evaluation board with a Zynq-7000 XC7Z045 system-on-chip is illustrated in \cite{bizimMDPI} to require approximately $11\%$ smaller hardware area and $64\%$ less processing time than the benchmark RO PUF hardware implementation in \cite{Pufky}.

%%%%%%%%%%%%%%%%%%%%%%%%%%%%%%%%%%%%%%%%%%%%%%%%%%%%%%%%%55
%            NIST Randomness Results
%%%%%%%%%%%%%%%%%%%%%%%%%%%%%%%%%%%%%%%%%%%%%%%%%%%%%%%%%
\subsection{Uniqueness and Security}\label{subsec:uniqueness}
The bit sequence extracted from a physical identifier should be uniformly distributed to make the rate region $\mathcal{R}_{\text{FCS}}$ in (\ref{eq:ls0}) valid. A common measure, called \textit{uniqueness}, for measuring randomness of bit sequences is the average fractional Hamming distance between the bit sequences extracted from different RO PUFs. Similar uniqueness results are obtained for all transforms, where the mean Hamming distance is $0.500$ and Hamming distance variance is approximately $\displaystyle 7\!\times \!10^{-4}$. All~transforms thus provide close to optimal uniqueness results due to their high decorrelation efficiencies and equipartitioned quantization intervals, which are better than previous RO PUF results with mean values of $0.462$ \cite{ROFirst} and $0.473$ \cite{ROLarge}.  

The national institute of standards and technology (NIST) provides a set of randomness tests that check whether a bit sequence can be differentiated from a uniformly random bit sequence \cite{NIST}. Apply these tests to measure the randomness of the generated sequences. The bit sequences generated from ROs in the dataset \cite{ROLarge} with the DWHT pass most of the applicable tests for short lengths, which is considered to be an acceptable result \cite{NIST}. The KLT performs the best due to its optimal decorrelation performance. One can apply a thresholding approach such that the reliable transform coefficients from which the bits are extracted do not have high correlations, which further improves the security performance. 

%%%%%%%%%%%%%%%%%%%%%%%%%%
% Error Correction
%%%%%%%%%%%%%%%%%%%%%%%%%
\section{Error-Correction Codes for PUFs with Transform Coding}\label{sec:correction}
Suppose that bit sequences extracted by using the transform-coding method are i.i.d. and uniformly distributed so that the secrecy leakage is zero. We assume that signal processing steps mentioned above perform well, so we can conduct standard information- and coding-theoretic analysis. We list different codes designed for the transform-coding algorithm according to the reliability metric considered above.

\begin{figure}
	\centering
	\newlength\figureheight
	\newlength\figurewidth
	\setlength\figureheight{5.2cm}
	\setlength\figurewidth{14.3cm}
\begin{tikzpicture}

\begin{axis}[%
width=0.951\figurewidth,
height=\figureheight,
at={(0\figurewidth,0\figureheight)},
scale only axis,
xmin=1,
xmax=10,
xlabel={Number of Quantization Bits $K$},
xmajorgrids,
ymin=0,
ymax=1,
ylabel={Correctness Probability $P_c$},
ymajorgrids,
axis background/.style={fill=white},
legend style={at={(0.0079,0.0107)},anchor=south west,legend cell align=left,align=left,draw=white!15!black}
]
\addplot [color=red,solid,line width=1.5pt,mark size=3.0pt,mark=asterisk,mark options={solid}]
  table[row sep=crcr]{%
1	0.998257301311316\\
2	0.995481003281785\\
3	0.990369641601798\\
4	0.98040809110094\\
5	0.960634304705393\\
6	0.92117005238719\\
7	0.842292962766348\\
8	0.690643212598228\\
9	0.481072154279916\\
10	0.297004683385764\\
11	0.17184382736417\\
12	0.0959793040700199\\
};
\addlegendentry{Coeff. 2};

\addplot [color=green,solid,line width=1.5pt,mark size=3.0pt,mark=o,mark options={solid}]
  table[row sep=crcr]{%
1	0.997829854852876\\
2	0.994372582050197\\
3	0.988007494814884\\
4	0.975602554847855\\
5	0.95097861539375\\
6	0.901834495690211\\
7	0.803718836699901\\
8	0.626612080739878\\
9	0.417072421387783\\
10	0.251075314644554\\
11	0.143338414886165\\
12	0.0794353605101738\\
};
\addlegendentry{Coeff. 1};

\addplot [color=blue,solid,line width=1.5pt,mark size=3.2pt,mark=square,mark options={solid}]
  table[row sep=crcr]{%
1	0.995957945354216\\
2	0.989518408888429\\
3	0.97766272562568\\
4	0.954557152546349\\
5	0.908692362891015\\
6	0.817204744400462\\
7	0.647964289458797\\
8	0.43737401312202\\
9	0.26530922001697\\
10	0.152081710529126\\
11	0.0844834747559681\\
12	0.0460313603210151\\
};
\addlegendentry{Coeff. 17};

\addplot [color=red,dashed,line width=1.5pt,mark size=5pt,mark=asterisk,mark options={solid}]
  table[row sep=crcr]{%
1	0.985222877647597\\
2	0.961673818771534\\
3	0.918315977543005\\
4	0.833830216915307\\
5	0.674630893567695\\
6	0.463852936793098\\
7	0.284268575193586\\
8	0.163836393340128\\
9	0.0913018064965708\\
10	0.0498483977069998\\
11	0.0268445209066746\\
12	0.0143139766736807\\
};
\addlegendentry{Coeff. 128};

\addplot [color=green,dashed,line width=1.5pt,mark size=3pt,mark=o,mark options={solid}]
  table[row sep=crcr]{%
1	0.984810457218178\\
2	0.960603711151823\\
3	0.916034816384811\\
4	0.829195710003641\\
5	0.66670380695252\\
6	0.455725042335783\\
7	0.278372640482159\\
8	0.160160315818596\\
9	0.0891633999024296\\
10	0.048649271931228\\
11	0.0261868464961171\\
12	0.013958574261824\\
};
\addlegendentry{Coeff. 150};

\addplot [color=blue,dashed,line width=1.5pt,mark size=2pt,mark=square,mark options={solid}]
  table[row sep=crcr]{%
1	0.984039637730627\\
2	0.95860356541513\\
3	0.911770979982468\\
4	0.820540284873643\\
5	0.652232492013706\\
6	0.441272707302553\\
7	0.268015323957184\\
8	0.153736674739541\\
9	0.0854365302130378\\
10	0.0465625576457378\\
11	0.0250434748870105\\
12	0.013341123560002\\
};
\addlegendentry{Coeff. 31};

\end{axis}
\end{tikzpicture}%
	\caption{The correctness probabilities for transform coefficients.} 
	\label{fig:correctnessprob}
\end{figure}
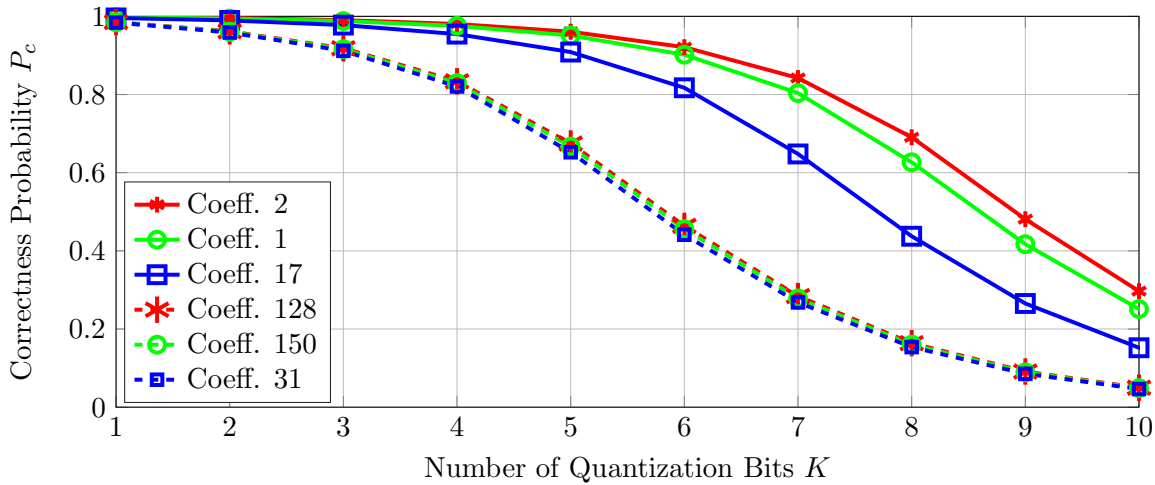

Select a channel code for the quantizer designed above for a fixed maximum number of errors to store a secret key of length 128 bits. The correctness probabilities defined in (\ref{eq:correctness}) for the transform coefficients with the three highest and three smallest probabilities are plotted in Figure~\ref{fig:correctnessprob}. The indices of the $16\times 16$ transform coefficients follow the order in the dataset \cite{ROLarge}, where the coefficient index at the first row and first column is $1$, and it increases columnwise up to $16$ so that the second row starts with the index $17$, the third row with the index $33$, etc. The most reliable transform coefficients are the low-frequency coefficients, which are in our case at the upper-left corner of the 2D transform-coefficient array with indices such as $1,2,3,17,18,19,33,34,$ and $35$. The low-frequency transform coefficients therefore have the highest signal-to-noise ratios (SNRs) for the source and noise statistics obtained from the RO dataset \cite{ROLarge}. The least reliable coefficients are observed to be spatially away from the transform coefficients at the upper-left or lower-right corners of the 2D transform-coefficient array. These results indicate that the \emph{SNR-packing efficiency}, which can be defined similarly as the energy-packing efficiency, of a transform follows a more complicated scan order than the classic zig-zag scan order used for the energy-packing efficiency. Observe~from Figure~\ref{fig:correctnessprob} that increasing the number of extracted bits decreases the correctness probability for all coefficients since the quantization boundaries get closer so that errors due to noise become more likely, {i.e.,} the probability $P_{c}(K)$ defined in (\ref{eq:correctness}) decreases with increasing $K$.

Fix the maximum number $\displaystyle C_{\text{max}}$ of transform coefficients allowed to be in error and calculate the correctness threshold $\displaystyle \thickbar{P}_c(C_{\text{max}})$ using (\ref{eq:threshold}), the total number $\displaystyle n(C_{\text{max}})$ of extracted bits using (\ref{eq:totalbits}), and the number $\displaystyle e(C_{\text{max}})$ of errors the block code should be able to correct using (\ref{eq:minimume}). Observe that if $\displaystyle C_{\text{max}}\!\leq\!10$, $\thickbar{P}_c(C_{\text{max}})$ is so large that $\displaystyle P_{c,i}(K\!=\!1)\!\leq\!\thickbar{P}_c(C_{\text{max}})$ for all $i=2,\ldots,l$. If $\displaystyle 11\!\leq\!C_{\text{max}}\!\leq\!15$, $\displaystyle n(C_{\text{max}})$ is less than the required code dimension of 128 bits. Furthermore, increasing $\displaystyle C_{\text{max}}$ results in a smaller correctness threshold $\displaystyle \thickbar{P}_c(C_{\text{max}})$ so that the maximum of the number $\displaystyle K_{\text{max}}(C_{\text{max}})\!=\!K^{\prime}_1(C_{\text{max}})$ of bits extracted among the $l-1$ used coefficients increases. This result can increase hardware complexity. Therefore, consider only the cases where $\displaystyle C_{\text{max}}\!\leq\!20$. Table~\ref{tab:myresults} shows $\displaystyle \thickbar{P}_c(C_{\text{max}})$, $\displaystyle n(C_{\text{max}})$, and $\displaystyle e(C_{\text{max}})$ for a range of $\displaystyle C_{\text{max}}$ values used for channel-code selection.

%%%%%%%%%
\begin{table}[t]
	\centering
	\caption{Code-parameter constraints.}
	\begin{tabular}{ |c|c|c|c|c|c| }
		\hline
		\rowcolor{blue!25}
		$\displaystyle \mathbf{C_{\text{\textbf{max}}}}$ & $\mathbf{16}$ & $\mathbf{17}$& $\mathbf{18}$ & $\mathbf{19}$ & $\mathbf{20}$\\
		\hline
		$\displaystyle \thickbar{P}_c$ &  $0.9902$
		&  $0.9889$  &   $0.9875$ &  $0.9860$ &  $0.9844$\\
		\hline
		$\displaystyle K_{\text{max}}$ &  $3$ &  $3$ &  $3$ & $3$ & $3$\\
		\hline
		$\displaystyle n$ & $144$ & $224$ &  $250$ & $255$ & $259$\\
		\hline
		$\displaystyle e$ &  $18$ &  $20$&  $21$ & $23$ & $25$\\
		\hline
	\end{tabular}\label{tab:myresults}
\end{table}

Consider binary (extended) Bose–Chaudhuri–Hocquenghem (BCH) and Reed-Solomon (RS) codes, which have good minimum-distance $d_{\text{min}}$ properties. An exhaustive search does not provide a code with dimension of at least 128 bits and with parameters satisfying any of the  $\displaystyle (n(C_{\text{max}}),\,e(C_{\text{max}}))$ pairs in Table~\ref{tab:myresults}. However, the correctness threshold analysis leading to Table~\ref{tab:myresults} is conservative. Thus, choose a BCH code with parameters as close as possible to an $\displaystyle (n(C_{\text{max}}),\,e(C_{\text{max}}))$ pair, for which one can prove that even if the number $\displaystyle e_{\text{BCH}}$ of errors the chosen BCH code can correct is less than $\displaystyle e(C_{\text{max}})$, the block-error probability constraint is satisfied. Consider the binary BCH code with the block length $255$, code dimension $131$, and a capability of correcting all error patterns with up to $\displaystyle e_{\text{BCH}}=18$ errors. 

%%%%%%%%%%%%%%%%%%%%%%%%%%%%%%%%%%%%%%%%%%%%%%

First, impose the condition that exactly one bit is extracted from each coefficient, {i.e.,} $K_{i}\!=\!1$ for all $i\!=\!2,3,\ldots,l$, so that in total $n\!=\!l-1\!=\!255$ bits are obtained, which results in mutually independent bit errors $E_i$. Therefore, the chosen block code should be able to correct all error patterns with up to $e\!=\!20$ bit errors rather than $e(20)\!=\!25$ bit errors, which is still greater than the error-correction capability $\displaystyle e_{\text{BCH}}=18$ of the considered BCH code.

The block error probability $P_B$ for the BCH code $\mathcal{C}(255,131,37)$ with a BMDD corresponds to the probability of having more than $18$ errors in the codeword, i.e., we obtain
\begin{align}
P_B = \sum_{j=19}^{255}\Bigg[\sum_{A\in\mathcal{F}_j}\prod_{i\in A}(1-P_{c,i})\,\bigcdot\prod_{i\in A^{c}}P_{c,i} \Bigg] \label{eq:blockerrorforbch}
\end{align}
where $P_{c,i}$ is the correctness probability of the $i$-th transform coefficient $\widehat{T}_i$ defined in (\ref{eq:correctness}) for $i\!=\!2,3,\ldots,256$, $\displaystyle \mathcal{F}_j$ is the set of all size-$j$ subsets of the set $\displaystyle\{2,3,\ldots,256\}$, and $A^{c}$ denotes the complement of the set $A$. The correctness probabilities $P_{c,i}$ are different and they represent probabilities of independent events due to the independence assumption for the transform coefficients. We use the discrete Fourier transform characteristic function method \cite{DFTCF} to calculate the block-error probability and obtain the result $P_B\!\approx\!1.26\!\times\!10^{-11}\!<\!10^{-9}$. The block-error probability constraint is thus satisfied by using the BCH code $\mathcal{C}(255,131,37)$ with a BMDD although the conservative analysis suggests that it would not be satisfied.

We next compare the BCH code $\mathcal{C}(255,131,37)$ with previous codes proposed for binding keys to physical identifiers with the FCS and a secret-key length of $128$ bits such that $P_B\!\leq\!10^{-9}$ is satisfied. The (secret-key, privacy-leakage) rate pair for this code is $(R_\text{s},R_\ell)=(\frac{131}{255},1\!-\!\frac{131}{255})\approx(0.514,\,0.486)$ bits/source-bit. This pair is significantly better than previous results. The main reason for obtaining a better (secret-key, privacy-leakage) rate pair is that the quantizer defined above allows to obtain a higher identifier-output reliability by decreasing the number of bits extracted from a transform coefficient.

\begin{figure}
	\centering
	\setlength\figureheight{5.75cm}
	\setlength\figurewidth{14.4cm}
	\input{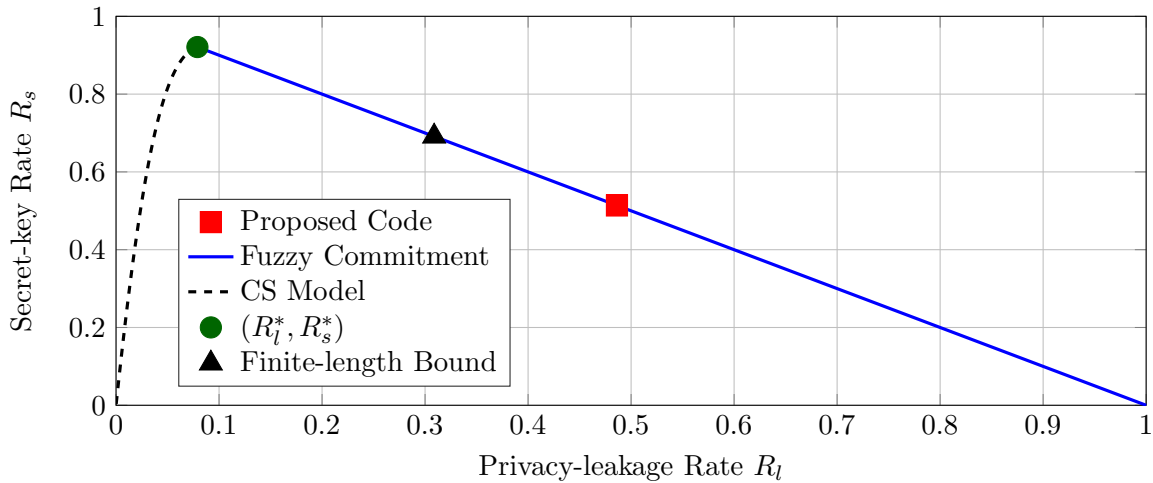}
	\caption{The operation point of the considered BCH code $\mathcal{C}(255,131,37)$, regions of achievable rate pairs according to (\ref{eq:ls0}) and (\ref{eq:chosensecret}), the maximum secret-key rate point $(R_\ell^*,R_{\text{s}}^*)$, and a finite-length bound for $n=255$ bits, $P_B=10^{-9}$, and BSC $(0.0097)$.} 
	\label{fig:ratecomparison}
\end{figure}

Compare the secret-key $R_\text{s}$ and privacy-leakage $R_\ell$ rates of the BCH code $\mathcal{C} (255,131,37)$ with the region of all achievable rate pairs for the CS model and the FCS for a BSC $P_{Y|X}$ with crossover probability $p_b\!=\! 1-\frac{1}{l-1}\sum_{i=2}^lP_{c,i}(K_i\!=\!1)\!\approx\!0.0097$, {i.e.,} the probability of being in error averaged over all used transform coefficients with the quantizer defined above. Compute the boundary points of the region $\mathcal{R}$ by finding the optimal auxiliary random variable $U$ in (\ref{eq:chosensecret}) when $P_{Y|X}$ is a BSC. The regions of all rate pairs achievable by the FCS and CS model, the maximum secret-key rate point, the (secret-key, privacy-leakage) rate pair of the BCH code, and a finite-length (non-asymptotic) bound \cite{Polyanskiy} for the block length of $n=255$ bits and $P_B\!=\!10^{-9}$ are plotted in Figure~\ref{fig:ratecomparison}.

%%%%%%%%%%%%%%%%%%%%%%%%%%%%%%%%%%%%%%%%%%%%%%%%%%%%%%%%

The maximum secret-key rate is $R_\text{s}^*\!\approx\!0.922$ bits/source-bit with a corresponding minimum privacy-leakage rate of $R_\ell^*\!\approx\!0.079$ bits/source-bit. There is a gap between the rate tuple achieved by the BCH code and the only operation point where the FCS is optimal, i.e., $(R_\ell^*,R_\text{s}^*)$. Part~of this rate loss can be explained by the short block length of the code and the small block-error probability constraint. The finite-length bound given in \cite[Theorem 52]{Polyanskiy} establishes that the rate pair $(R_\text{s},R_\ell)\!=\!(0.691,0.309)$ bits/source-bit is achievable by using the FCS, as depicted in Figure~\ref{fig:ratecomparison}. One can therefore further improve the rate pairs by using better channel codes and decoders with higher hardware complexity, but this may not be possible for IoT applications. Figure~\ref{fig:ratecomparison} also illustrates that there exist other code constructions (other than standard channel codes) that~reduce the privacy-leakage rate as well as the storage rate for each fixed secret-key rate, which we consider below.

%%%%%%%%%%%%%%%%%%%%%%%%%%%%%%%%%%%%%%%%%%%%%%
%%%%%%%%%%%%%%%%%%%%%%%%%%%%%%%%%%%%%%%%%%%%%%
%%%%%%%%%%%%%%%%%%%%%%%%%%%%%%%%%%%%%%%%%%%%%%
%%%%%%%%%%%%%%%%%%%%%%%%%%%%%%%%%%%%%%%%%%%%%%
\section{Code Constructions for PUFs}\label{sec:WZchapter}
Consider the two-terminal key agreement problem, where the identifier outputs during enrollment are noiseless. We mention two optimal linear code constructions from \cite{bizimWZ} that are based on Wyner-Ziv (WZ) coding \cite{WZCard}. The first construction uses random linear codes and achieves all points of the key-leakage-storage regions of the GS and CS models. The second construction uses nested polar codes for vector quantization during enrollment and for error correction during reconstruction. Simulations show that nested polar codes achieve privacy-leakage and storage rates that improve on existing code designs, and one designed code achieves a rate tuple that cannot be achieved by existing methods. 

Several practical code constructions for key agreement with identifiers have been proposed in the literature. For instance, the COFE and the FCS both require a standard error-correction code to satisfy the constraints of, respectively, the key generation (GS model) and key embedding (CS model) problems, as discussed above. Similarly, a polar code construction is proposed in \cite{IgnaPolar} for the GS model. These constructions are shown to be suboptimal in terms of the privacy-leakage and storage rates. 

The binary Golay code is used in \cite{IgnaTrans} as a vector quantizer (VQ) in combination with Slepian-Wolf (SW) codes \cite{SW} to illustrate that the key vs. storage (or key vs. leakage) rate ratio can be increased via quantization. This observation motivates the use of a VQ to improve the performance of previous constructions. We next consider VQ by using WZ coding to decrease storage rates. The WZ-coding construction turns out to be optimal, which is not coincidental. For instance, the bounds on the storage rate of the GS model and on the WZ rate (storage rate) have the same mutual information terms optimized over the same conditional probability distribution. This similarity suggests an \emph{equivalence} that is closely related to the concept of \emph{formula duality}. In fact, the optimal random code construction, encoding, and decoding operations are identical for both problems. One therefore can call the GS model and WZ problem \emph{functionally equivalent}. Such a strong connection suggests that there might exist constructive methods that are optimal for both problems for all channels, which is closely related to the \emph{operational duality} concept.

%%%%%%%%%%%%%%%%%%%%%%%%%%%%%%%%%%%%%%%%%%%%%%%%%%
\begin{figure}
	\centering
	\resizebox{0.61\linewidth}{!}{
		
		\begin{tikzpicture}
		% Source Box
		\node (so) at (-1.5,-2.2) [draw,rounded corners = 5pt, minimum width=1.0cm,minimum height=0.8cm, align=left] {$P_X$};
		% Enc Box
		\node (a) at (0,0) [draw,rounded corners = 6pt, minimum width=3.2cm,minimum height=1.2cm, align=left] {$
			(S,W) \overset{(a)}{=} \Enc(X^n)$\\ $W\overset{(b)}{=}\Enc(X^n,S)$};
		% Channel Box
		\node (c) at (3,-2.2) [draw,rounded corners = 5pt, minimum width=1.6cm,minimum height=0.8cm, align=left] {$P_{Y|X}$};
		% Dec Box
		\node (b) at (6,0) [draw,rounded corners = 6pt, minimum width=3.2cm,minimum height=1.2cm, align=left] {$\widehat{S} = \Dec\left(Y^n,W\right)$};
		% Enc to Dec Arrow
		\draw[decoration={markings,mark=at position 1 with {\arrow[scale=1.5]{latex}}},
		postaction={decorate}, thick, shorten >=1.4pt] (a.east) -- (b.west) node [midway, above] {$W$};
		% X^N goes to Enc Box
		\node (a1) [below of = a, node distance = 2.2cm] {$X^n$};
		\node (b1) [below of = b, node distance = 2.2cm] {$Y^n$};
		\node (k9) [below of = a1, node distance = 0.6cm] {Enrollment};
		\node (k19) [below of = b1, node distance = 0.6cm] {Reconstruction};
		% Source to X^N arrow
		\draw[decoration={markings,mark=at position 1 with {\arrow[scale=1.5]{latex}}},
		postaction={decorate}, thick, shorten >=1.4pt] (so.east) -- (a1.west);
		% X^N  to Enc arrow
		\draw[decoration={markings,mark=at position 1 with {\arrow[scale=1.5]{latex}}},
		postaction={decorate}, thick, shorten >=1.4pt] (a1.north) -- (a.south);
		% X^N  to Channel box arrow
		\draw[decoration={markings,mark=at position 1 with {\arrow[scale=1.5]{latex}}},
		postaction={decorate}, thick, shorten >=1.4pt] (a1.east) -- (c.west);
		% Channel box to Y^N arrow
		\draw[decoration={markings,mark=at position 1 with {\arrow[scale=1.5]{latex}}},
		postaction={decorate}, thick, shorten >=1.4pt] (c.east) -- (b1.west);
		% Y^N  to Enc arrow
		\draw[decoration={markings,mark=at position 1 with {\arrow[scale=1.5]{latex}}},
		postaction={decorate}, thick, shorten >=1.4pt] (b1.north) -- (b.south);
		% S exchanges with Enc
		\node (a2) [above of = a, node distance = 2.2cm] {$\;S$};
		% \hat{S} exchanges with Dec
		\node (b2) [above of = b, node distance = 2.2cm] {$\;\widehat{S}$};
		% Dec to \hat{S} arrow
		\draw[decoration={markings,mark=at position 1 with {\arrow[scale=1.5]{latex}}},
		postaction={decorate}, thick, shorten >=1.4pt] ($(b.north)-(0.3,0)$) -- ($(b2.south)-(0.3,0)$) node [midway, left] {$(a)$};
		% Dec to \hat{S}' arrow
		\draw[decoration={markings,mark=at position 1 with {\arrow[scale=1.5]{latex}}},
		postaction={decorate}, thick, shorten >=1.4pt] ($(b.north)+(0.3,0)$)-- ($(b2.south)+(0.3,0)$) node [midway, right] {$(b)$};
		% Enc to S arrows
		\draw[decoration={markings,mark=at position 1 with {\arrow[scale=1.5]{latex}}},
		postaction={decorate}, thick, shorten >=1.4pt] ($(a.north)-(0.3,0)$)-- ($(a2.south)-(0.3,0)$) node [midway, left] {$(a)$};
		% S to Enc arrow
		\draw[decoration={markings,mark=at position 1 with {\arrow[scale=1.5]{latex}}},
		postaction={decorate}, thick, shorten >=1.4pt]  ($(a2.south)+(0.3,0)$)-- ($(a.north)+(0.3,0)$) node [midway, right] {$(b)$};
		\end{tikzpicture}
	}
	\caption{The $(a)$ GS and $(b)$ CS models.}\label{fig:problemsetup}
\end{figure}
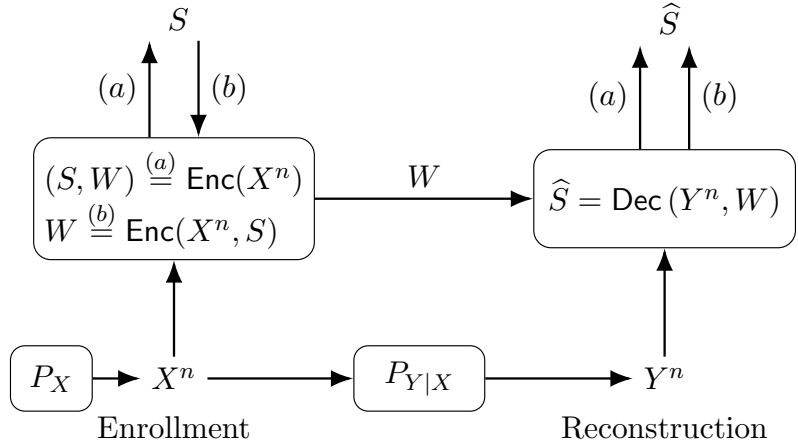
%%%%%%%%%%%%%%%%%%%%%%%%%%%%%%%%%%%%%%%%%%%%%%%%%%%%%%%%%%%%%%%%%

Consider the GS model in Figure~\ref{fig:problemsetup}(a), where a secret key is generated from a biometric or physical source. During enrollment, the encoder observes an i.i.d. noiseless sequence $X^n$, generated by the source according to some $P_X$, and computes a secret key $S$ and public helper data $W$ as $\displaystyle (S,W)\,{=}\,{\Enc}(X^n)$. During reconstruction, the decoder observes a noisy source measurement $Y^n$ of the source $X^n$ through a memoryless channel $P_{Y|X}$ together with the helper data $W$. The decoder estimates the secret key as $\displaystyle \widehat{S}\,{=}\,{\Dec}(Y^n\!,W)$. Similarly, Figure~\ref{fig:problemsetup}(b) shows the CS model, where a secret key $S$ that is independent of $(X^n,Y^n)$ is embedded into the helper data as $W = \Enc(X^n,S)$. The decoder for the CS model estimates the secret key as $\widehat{S}=\Dec(Y^n,W)$. The source, measurement, secret key, and storage alphabets $\mathcal{X}$, $\mathcal{Y}$, $\mathcal{S}$, and $\mathcal{W}$ are finite sets, which can be achieved if, e.g., the transform-coding algorithm discussed above is applied.

\begin{definition}\label{def:achievabilityGSCS}
	A key-leakage-storage tuple $(R_\text{s},R_\ell,R_\text{w})$ is \emph{achievable} for GS and CS models if, given any $\delta>0$, there is some $n\!\geq\!1$, an encoder, and a decoder such that $\displaystyle R_\text{s}=\frac{\log|\mathcal{S}|}{n}$ and 
	\begin{alignat}{2}
	&P_B=\Pr[S\ne\hat{S}] \leq \delta &&\qquad\quad (reliability) \label{eq:reliabilityconstMMMM}\\
	&I\left(S;W\right) \leq n\delta &&\qquad\quad(weak\; secrecy)  \label{eq:secrecyconstMMMM}\\
	&I\left(X^n;W\right) \leq n(R_\ell+\delta) \quad\quad\quad&&\qquad\quad (privacy)  \label{eq:privacyconstMMMM}\\
	&H(S) \geq n(R_\text{s}-\delta)  \quad&&\qquad\quad (uniformity)\label{eq:uniformityconstMMMM}\\
	&\log|\mathcal{W}| \le n(R_\text{w}+\delta)  &&\qquad\quad (storage)\label{eq:storageconstMMMM}
	\end{alignat}
	are satisfied. The \emph{key-leakage-storage} regions $\mathcal{R}_{\text{gs}}$ and $\mathcal{R}_{\text{cs}}$ for the GS and CS models, respectively, are the closures of the sets of achievable tuples for the corresponding models.
\end{definition}

\begin{theorem}[\hspace{1sp}\cite{IgnaTrans}]\label{theo:secrecyregions}
	The key-leakage-storage regions for the GS and CS models, respectively, are 
	\begin{align}
	\begin{split}
	\mathcal{R}_{\text{gs}}\! =\! &\bigcup_{P_{U|X}}\!\Big\{\left(R_\text{s},R_\ell,R_\text{w}\right)\!\colon\! \nonumber\\
	&0\leq R_\text{s}\leq I(U;Y),\nonumber\\
	&R_\ell\geq I(U;X)-I(U;Y),\nonumber\\
	&R_\text{w}\geq I(U;X)-I(U;Y)\Big\} \text{,}%\label{eq:regionGS}
	\end{split}
	\qquad\quad\text{ and }\qquad
	\begin{split}
	\mathcal{R}_{\text{cs}}\! =\! &\bigcup_{P_{U|X}}\!\Big\{\left(R_\text{s},R_\ell,R_\text{w}\right)\!\colon\! \nonumber\\
	&0\leq R_\text{s}\leq I(U;Y),\nonumber\\
	&R_\ell\geq I(U;X)-I(U;Y),\nonumber\\
	&R_\text{w}\geq I(U;X)\Big\} %\label{eq:regionCS}
	\end{split}
	\end{align}
	where $U-X-Y$ form a Markov chain. These regions are convex sets. The alphabet $\mathcal{U}$ of the auxiliary random variable $U$ can be limited to have size $\displaystyle |\mathcal{U}|\!\leq\!|\mathcal{X}|+1$ for both regions.
\end{theorem}

\begin{remark}
	One can improve the weak secrecy to strong secrecy, i.e., we can replace (\ref{eq:secrecyconstMMMM}) with $I(S;W)\leq \delta$ by applying information reconciliation and privacy amplification steps to multiple blocks of identifier outputs as described in \cite{MaurerSecrecyFree}, e.g., by using multiple PUFs in a device for key agreement. 	
\end{remark}

Assume, as above, that $X^n\sim \text{Bern}^n(\frac{1}{2})$ and the channel $P_{Y|X}$ is a BSC$(p_A)$, where $p_A\in[0, 0.5]$. Define the star-operation as $q*p_A = q(1-p_A)+(1-q)p_A$. The key-leakage-storage region of the GS model is
\begin{align}
\mathcal{R}_{\text{gs},\text{bin}}\! =\! &\bigcup_{q\in[0,0.5]}\!\Big\{\left(R_\text{s},R_\ell,R_\text{w}\right)\!\colon\!\nonumber\\
&0\leq R_\text{s}\leq 1- H_b(q*p_A),\nonumber\\
&R_\ell\geq H_b(q*p_A)- H_b(q),\nonumber\\
&R_\text{w}\geq H_b(q*p_A)- H_b(q)\Big\}\label{eq:BSCRegionGS}.
\end{align}

%%%%%%%%%%%%%%%%%%%%%%%%%%%%%%%%%%%%%%%%%%%%%%%%%%%%%%%%%%%%%%%%%%%%%%%%%%%%%%%%%%%%%%
%%%%%%%%%%%%%%%%%%%%%%%%%%%%%%%%%%%%%%%%%%%%%%%%%%%%%%%%%%%%%%%%%%%%%%%%%%%%%%%%%%%%%%
%%%%%%%%%%%%%%%%%%%%%%%%%%%%%%%%%%%%%%%%%%%%%%%%%%%%%%%%%%%%%%%%%%%%%%%%%%%%%%%%%%%%%%
\subsection{Comparisons Between Code Constructions for PUFs}\label{subsec:CompareMethods}
There are several existing code constructions proposed for the GS and CS models. Consider the three best methods: FCS for the CS model, and COFE and the polar code construction in \cite{IgnaPolar} for the GS model, to compare them with the WZ-coding constructions. Similar steps to the FCS are applied in the COFE, except that the secret key is a hashed version of $X^n$. The FCS achieves the single optimal point in the key-leakage region with the maximum secret-key rate $R_\text{s}^*=I(X;Y)$; the privacy-leakage rate is $R_\ell^* =H(X|Y)$. Similarly, the COFE achieves the same boundary point in the key-leakage region. This is, however, the only boundary point of the key-leakage regions that these methods can achieve.

One can improve both methods by adding a VQ step: instead of $X^n$ we use its quantized version $X_q^n$ during enrollment. This asymptotically corresponds to summing the original helper data and an independent random variable $J^n\sim\text{Bern}^n(q)$ such that $W=X^n\xor C^n\xor J^n$ is the new helper data so that we create a virtual channel $P_{Y|X\xor J}$ and apply the FCS or COFE to this virtual channel. The modified FCS and COFE can achieve all points of the key-leakage region if we take a union of all rate pairs achieved over all $q\in[0, 0.5]$. However, the helper data have $n$ bits for both methods, and the resulting storage rate of $1$ bit/source-bit is not necessarily optimal.

The polar code construction in \cite{IgnaPolar} requires less storage rate than the FCS and COFE. However, this approach improves only the storage rate and cannot achieve all points of the key-leakage-storage region. Furthermore, in \cite{IgnaPolar} some code designs assume that there is a ``private'' key shared only between the encoder and decoder, which is not realistic since a private key requires hardware protection against invasive attacks. If such a protection is possible, then there is no need to use an on-demand key reconstruction method like a PUF. 

The existing methods cannot, therefore, achieve all points of the key-leakage-storage region for a BSC, unlike the WZ-coding constructions described in \cite{bizimWZ} and illustrated with nested polar code designs below. In previous works, only the secret-key rates of the proposed codes are compared because the sum of the secret-key and privacy-leakage rates is one. This constraint means that increasing the key vs. leakage (or key vs. storage) rate ratio is equivalent to increasing the key rate. Instead, WZ-coding constructions are more flexible than the existing methods in terms of achievable rate tuples. Therefore, use the key vs. storage rate ratio as the metric to control the storage and privacy leakage.

%%%%%%%%%%%%%%%%%%%%%%%%%%%%%%%%%%%%%%%%%%%%%%%%%%%%%%%%%%%%%%%%%%%%%%%%%%%%%%%%%%%%%%
\section{Optimal Code Constructions with Polar Codes}\label{sec:codeproposal}
Polar codes \cite{Arikan} have a low encoding/decoding complexity, asymptotic optimality for various information-theoretic problems, and good finite length performance if a list decoder is used in combination with an outer code. Furthermore, they have a structure that allows a simple nested code design and they can be used for WZ coding \cite{RudigerPolarExtended}. 

Polar codes rely on the \textit{channel polarization} phenomenon, where a channel is converted into polarized bit channels by a polar transform. This transform converts an input sequence $U^n$ with frozen and unfrozen bits to a codeword of the same length $n$. A polar decoder processes a noisy observation of the codeword together with the frozen bits to estimate ${U}^n$.

Let $\mathcal{C}(n,\mathcal{F},G^{|\mathcal{F}|})$ denote a polar code of block length $n$, where $\mathcal{F}$ is the set of indices of the frozen bits and $G^{|\mathcal{F}|}$ is the sequence of frozen bits. In the following, we use the nested polar code construction proposed in \cite{RudigerPolarExtended} for WZ coding.

%%%%%%%%%%%%%%%%%%%%%%%%%%%%%%%%%%%%%
\subsection{Polar Code Construction for the GS Model}\label{subsec:polarcons}
Consider two polar codes $\mathcal{C}_1(n,\mathcal{F}_1, V)$ and $\mathcal{C}(n,\mathcal{F}, \xoverline{V})$ with $\mathcal{F}=\mathcal{F}_1 \cup \mathcal{F}_w$ and $\xoverline{V}=[V, W]$, where $V$ has length $m_1$ and $W$ has length $m_2$ such that $m_1$ and $m_2$ satisfy
\begin{align}
\begin{split}
&\frac{m_1}{n} = H_b(q)-\delta%\label{eq:constraintonm1}
\end{split}
\qquad\qquad\quad\text{ and }\quad
\begin{split}
&\frac{m_1+m_2}{n} = H_b(q*p_A)+\delta%\label{eq:constraintonm2}
\end{split}
\end{align}   
for some distortion $q\in[0,0.5]$ and $\delta>0$. The indices in $\mathcal{F}_1$ represent frozen channels with assigned values $V$ for both codes and $\mathcal{C}$ has additional frozen channels with assigned values $W$ denoted by the set of indices $\mathcal{F}_w$, so the codes are nested. 

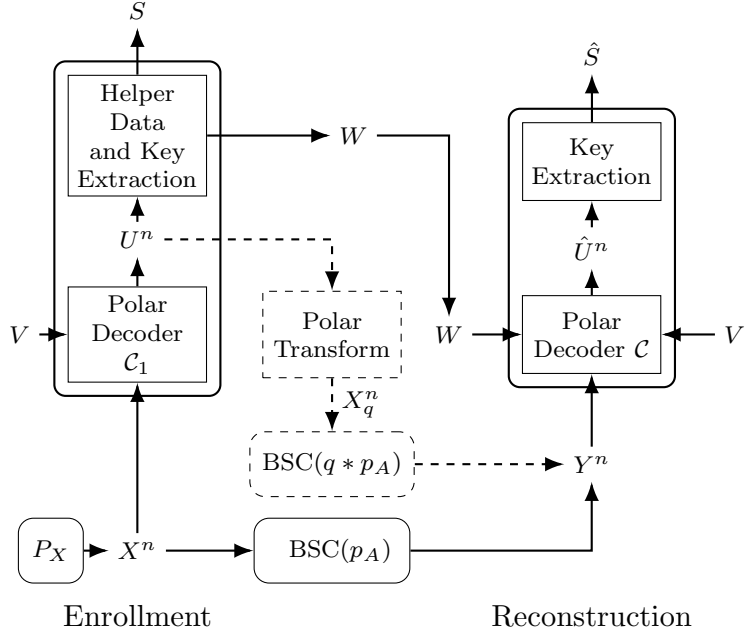
\begin{figure}
	\centering
	\begin{tikzpicture}[auto,>=latex', scale = 1.15, transform shape]
\centering
\scriptsize
\node[draw, rectangle,align=center, minimum height = 0.75cm, minimum width= 0.75 cm,rounded corners=.2cm](PUF){$P_X$};
\node[right of =PUF,node distance = 1cm](X){$X^n$};

\node[draw, right of=X,rectangle,align=center, node distance =2.25cm, minimum height = 0.75cm, minimum width= 1.8 cm, text width = 1.0cm,rounded corners=.2cm] (PYX) {$\text{BSC}(p_A)$};

\node[right of =PYX,node distance = 3cm](Y){};

\node[above of =Y,node distance = 1cm](Y2){$Y^n$};

\node[draw, above of=X,rectangle,align=center, node distance =2.5cm, minimum height = 0.9cm, minimum width= 1.4 cm, text width = 1.4cm] (VQ) {Polar Decoder $\mathcal{C}_1$};

\node[above of =VQ,node distance = 1.1cm](U){$U^n$};

\node[draw, above of=U,rectangle,align=center, node distance =1.2cm, minimum height = 0.85cm, minimum width= 1.4 cm, text width = 1.4cm] (SExt) {Helper Data and Key Extraction};

\node[above of =SExt,node distance = 1.45cm](S){$S$};

\node[right of =SExt,node distance = 2.5cm](W){$W$};

\node[draw, above of=Y,rectangle,align=center,node distance =2.5cm, minimum height = 0.9cm, minimum width= 1.4 cm, text width = 1.4cm] (Dec) {Polar Decoder $\mathcal{C}$};

\node[above of =Dec,node distance = 1cm](Uh){$\hat{U}^n$};

\node[draw, above of=Uh,rectangle,align=center, node distance =1cm, minimum height = 0.9cm, minimum width= 1.4 cm, text width = 1.4cm] (RExt) {Key Extraction};

\node[above of =RExt,node distance = 1.25cm](Sh){$\hat{S}$};

\node[left of =VQ,node distance = 1.36cm](V){$V$};
\node[left of =Dec,node distance = 1.65cm](VW){$W$};
\node[right of =Dec,node distance = 1.65cm](WV){$V$};

\node[draw, right of=VQ,rectangle,align=center, node distance =2.25cm, minimum height = 1cm, minimum width= 1.4 cm, text width = 1.4cm,dashed] (Enc) {Polar Transform};

\node[draw, above of=PYX,rectangle,align=center, node distance =1cm, minimum height = 0.75cm, minimum width= 1.8 cm, text width = 1.7cm,rounded corners=.2cm,dashed] (PXq) {BSC$(q*p_A)$};

\draw[decoration={markings,mark=at position 1 with {\arrow[scale=1.4]{latex}}},
postaction={decorate}, thick, shorten >=1.4pt] (PUF) -- (X); 
\draw[decoration={markings,mark=at position 1 with {\arrow[scale=1.4]{latex}}},
postaction={decorate}, thick, shorten >=1.4pt] (X) -- (PYX); 
\draw[decoration={markings,mark=at position 1 with {\arrow[scale=1.4]{latex}}},
postaction={decorate}, thick, shorten >=1.4pt] (PYX) -| (Y2);
\draw[decoration={markings,mark=at position 1 with {\arrow[scale=1.4]{latex}}},
postaction={decorate}, thick, shorten >=1.4pt] (X) -- (VQ);
\draw[decoration={markings,mark=at position 1 with {\arrow[scale=1.4]{latex}}},
postaction={decorate}, thick, shorten >=1.4pt] (VQ) -- (U);
\draw[decoration={markings,mark=at position 1 with {\arrow[scale=1.4]{latex}}},
postaction={decorate}, thick, shorten >=1.4pt] (U) -- (SExt);
\draw[decoration={markings,mark=at position 1 with {\arrow[scale=1.4]{latex}}},
postaction={decorate}, thick, shorten >=1.4pt] (SExt) -- (S);
\draw[decoration={markings,mark=at position 1 with {\arrow[scale=1.4]{latex}}},
postaction={decorate}, thick, shorten >=1.4pt] (SExt) -- (W);

\draw[decoration={markings,mark=at position 1 with {\arrow[scale=1.4]{latex}}},
postaction={decorate}, thick, shorten >=1.4pt] (Y2) -- (Dec);
\draw[decoration={markings,mark=at position 1 with {\arrow[scale=1.4]{latex}}},
postaction={decorate}, thick, shorten >=1.4pt] (Dec) -- (Uh);
\draw[decoration={markings,mark=at position 1 with {\arrow[scale=1.4]{latex}}},
postaction={decorate}, thick, shorten >=1.4pt] (Uh) -- (RExt);
\draw[decoration={markings,mark=at position 1 with {\arrow[scale=1.4]{latex}}},
postaction={decorate}, thick, shorten >=1.4pt] (RExt) -- (Sh);

\draw[decoration={markings,mark=at position 1 with {\arrow[scale=1.4]{latex}}},
postaction={decorate}, thick, shorten >=1.4pt] (V) -- (VQ);
\draw[decoration={markings,mark=at position 1 with {\arrow[scale=1.4]{latex}}},
postaction={decorate}, thick, shorten >=1.4pt] (VW) -- (Dec);
\draw[decoration={markings,mark=at position 1 with {\arrow[scale=1.4]{latex}}},
postaction={decorate}, thick, shorten >=1.4pt] (WV) -- (Dec);
\draw[decoration={markings,mark=at position 1 with {\arrow[scale=1.4]{latex}}},
postaction={decorate}, thick, shorten >=1.4pt] (W) -| (VW.north);

\draw[decoration={markings,mark=at position 1 with {\arrow[scale=1.4]{latex}}},
postaction={decorate}, thick, shorten >=1.4pt,dashed] (U) -| (Enc);

\draw[decoration={markings,mark=at position 1 with {\arrow[scale=1.4]{latex}}},
postaction={decorate}, thick, shorten >=1.4pt,dashed] (Enc) -- (PXq) node[midway] {$X_q^n$};
\draw[decoration={markings,mark=at position 1 with {\arrow[scale=1.4]{latex}}},
postaction={decorate}, thick, shorten >=1.4pt,dashed] (PXq) -- (Y2);

\node (redrect1) [thick,rectangle, draw, fit=(Dec)(RExt),inner sep=5pt,color=black,solid,rounded corners=.15cm] {};
\node (redrect2) [thick,rectangle, draw, fit=(VQ)(SExt),inner sep=5pt,color=black,solid,rounded corners=.15cm] {};
\small
\node[below of =X,node distance = 0.75cm](Enr){Enrollment};
\node[below of =Y,node distance = 0.75cm](Rec){Reconstruction};

\end{tikzpicture}
	\caption{Second WZ-coding construction for the GS model.}
	\label{fig:blockdig}
\end{figure}

The code $\mathcal{C}_1$ serves as a VQ with a desired distortion $q$ since its rate is greater than the lossy source coding capacity with average distortion $q$. The code $\mathcal{C}$ serves as the error-correction code for a BSC($q*p_A$) since its rate is less than channel capacity of this channel. The idea is to obtain $W$ during enrollment and store it as public helper data. For reconstruction, $(W,V,Y^n)$ are used by the decoder to estimate the secret key $S$ of length $n-m_1-m_2$. Figure~\ref{fig:blockdig} shows the block diagram of this construction. Suppose $V$ is the all-zero vector so that no additional storage is necessary. This choice has no effect on the average distortion $E[q]$ between $X^n$ and $X_q^n$ defined below; see \cite[Lemma 10]{RudigerPolarExtended}.

\textit{Enrollment}: The uniform binary sequence $X^n$ generated by a PUF during enrollment is treated as the noisy observation of a BSC$(q)$. $X^n$ is quantized by a polar decoder of $\mathcal{C}_1$. Extract from the decoder output $U^n$ the bits at indices $\mathcal{F}_w$ and store them as the helper data $W$. The bits at the indices $j\in \{1,2,\ldots,n\}\setminus \mathcal{F}$ are used as the secret key. Note that applying the polar transform to $U^n$ generates $X_q^n$, which is a distorted version  of $X^n$. The distortion between $X^n$ and $X_q^n$ is modeled as a BSC($q$) because the error sequence $E_{q}^n=X^n\xor X_q^n$ resembles an i.i.d. sequence $\sim\text{Bern}^n(q)$ when $n\rightarrow \infty$ \cite[Lemma 11]{RudigerPolarExtended}. 

\textit{Reconstruction}: During reconstruction, the polar decoder of $\mathcal{C}$ observes the binary sequence $Y^n$, which is a noisy measurement of $X^n$ through a BSC$(p_A)$. The frozen bits $\xoverline{V}=[V, W]$ at indices $\mathcal{F}$ are given to the polar decoder. The output $\widehat{U}^n$ of the polar decoder is the estimate of $U^n$ and contains the estimate $\widehat{S}$ of the secret key at the unfrozen indices of $\mathcal{C}$, i.e., $j\in \{1,2,\ldots,n\}\setminus \mathcal{F}$.

We next summarise a method to design practical nested polar codes for the GS model.

\textit{Construction of $\mathcal{C}$ and $\mathcal{C}_1$}: Since $\mathcal{C}\subseteq\mathcal{C}_1$ are nested codes, they must be constructed jointly. $\mathcal{F}$ and $\mathcal{F}_{1}$ should be chosen such that the reliability and security constraints are satisfied. For a given secret key size $n-m_1-m_2$, block length $n$, crossover probability $p_A$, and target block-error probability $P_B=\Pr[S\ne\widehat{S}]$, consider the following nested polar code design procedure \cite{bizimWZ}.
\begin{enumerate}
	\item Construct a polar code of rate $(n\!-\!m_1\!-\!m_2)/n$ and use it as the code $\mathcal{C}$, i.e., define the set of frozen indices $\mathcal{F}$.
	\item Evaluate the error correction performance of $\mathcal{C}$ with a decoder for a BSC over a range of crossover probabilities to obtain the crossover probability $p_c$, resulting in a target block-error probability of $P_B$. Using $p_c=E[q]*p_A$, we obtain the target distortion $E[q]=(p_c-p_A)/(1-2p_A)$ averaged over a large number of realizations of $X^n$.
	\item Find an $\mathcal{F}_1\subset \mathcal{F}$ that results in an average distortion of $E[q]$ with a minimum possible amount of helper data. Use $\mathcal{F}_1$ as the frozen set of $\mathcal{C}_1$. 
\end{enumerate}
Step 1 is a conventional channel code design task, similar to the codes designed for the transform-coding algorithm above, and step 2 is applied by Monte-Carlo simulations. For step 3, we start with  $\mathcal{F}_1^{'} = \mathcal{F}$ and compute the resulting average distortion $E[q']$ via Monte-Carlo simulations. If $E[q']$ is not less than $E[q]$, remove elements from $\mathcal{F}_1^{'}$ according to the reliabilities of the polarized bit channels and repeat the procedure until we obtain the desired average distortion $E[q]$. 

The distortion level introduced by the VQ is an additional degree of freedom in choosing the code design parameters. For instance, different values of $P_B$ can be targeted with the same code by changing the distortion level. Alternatively, devices with different $p_A$ values can be supported by using the same code. This additional degree of freedom makes the mentioned code design suitable for a wide range of applications.

%%%%%%%%%%%%%%%%%%%%%%%%%%%%%%%%%%%%%
\subsection{Designed Polar Codes for the GS Model}
Consider, e.g., the GS model where $S$ is used in the AES and $\log|\mathcal{S}|\!=\!n\!-\!m_1\!-\!m_2\!=\!128$ bits, as considered above. If we use PUFs in an FPGA as the randomness source, we must satisfy a block-error probability $P_B$ of at most $10^{-6}$. Consider a BSC $P_{Y|X}$ with crossover probability $p_A=0.15$, which is a common value for SRAM PUFs under ideal environmental conditions \cite{SoftHelper} and for RO PUFs under varying environmental conditions \cite{bizimtemperature}. Consider nested polar codes for these parameters to illustrate that one can achieve better key-leakage-storage rate tuples than previously proposed codes. 

\emph{Code 1}: Consider $n=1024$ bits and recall that $n-m_1-m_2=128$ bits, $P_B=10^{-6}$, and $p_A=0.15$. Polar successive cancellation list (SCL) decoders with list size $8$ are used as the VQ and channel decoder. First, design the code $\mathcal{C}$ of rate $128/1024$ and evaluate its performance with the SCL decoder for a BSC with a range of crossover probabilities. A block-error probability of $10^{-6}$ is observed at a crossover probability of $p_c=0.1819$. Since $p_A=0.15$, this corresponds to an average distortion of $E[q]=0.0456$ and the target average distortion is obtained at $n-m_1=778$ bits. Thus, $m_2=650$ bits of helper data suffice to obtain a block-error probability of $P_B=10^{-6}$.

\emph{Code 2}: Consider the same parameters as in Code 1, except $n=2048$ bits. Apply the same steps as above. A crossover probability of $p_c=0.2682$ is required to obtain a block-error probability of $10^{-6}$, which gives an average distortion of $E[q]=0.1689$. The target average distortion is achieved with helper data of length $611$ bits. 

The error probability $P_B$ is calculated as an average over a large number of PUF realizations, i.e., over a large number of PUF devices with the same circuit design. To satisfy the block-error probability requirement for each PUF realization, one could consider using the maximum distortion instead of $E[q]$ in step 3 of the design procedure given above. This would increase the amount of helper data. One can guarantee for the considered parameters a block-error probability of at most $10^{-6}$ for $99.99\%$ of all realizations $x^n$ of $X^n$ by adding $32$ bits to the helper data for Code 1 and $33$ bits for Code 2. The numbers of extra helper data bits required are small since the variance of the distortion $q$ over all PUF realizations is small for the blocklengths considered. For comparisons, we consider the helper data sizes required to guarantee $P_B=10^{-6}$ for $99.99\%$ of all PUF realizations.

\begin{figure*}
	\centering
	\setlength\figureheight{5.75cm}
	\setlength\figurewidth{16.4cm}
	\input{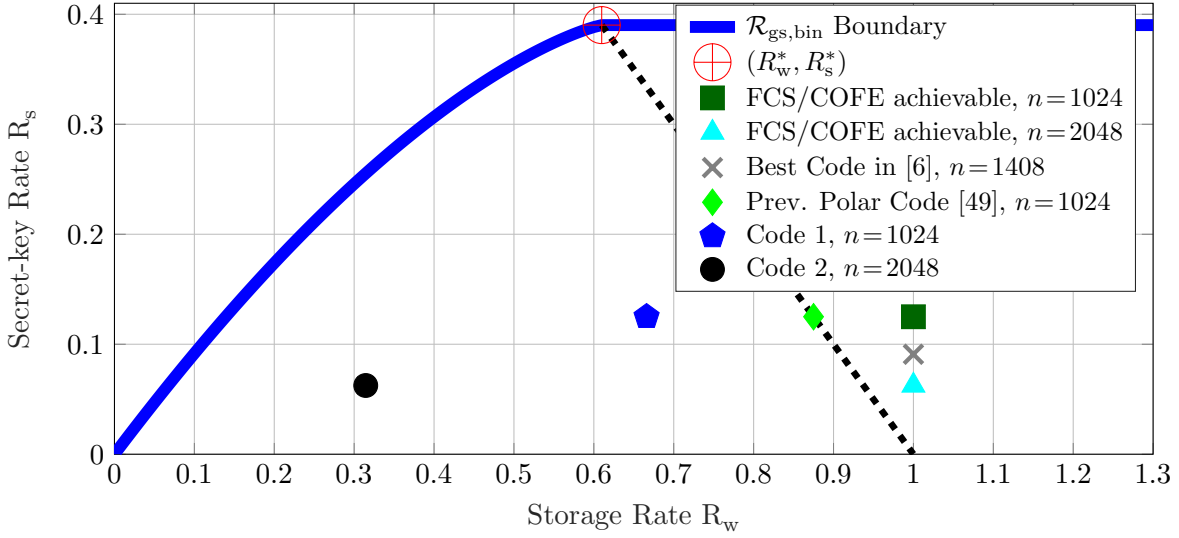}
	\caption{Storage-key rates for the GS model with $p_A=0.15$. The $(R_\text{w}^*,R_\text{s}^*)$ point is the best possible point achieved by SW-coding constructions, which lies on the dashed line representing $R_\text{w}+R_\text{s} = H(X)$. The block error probability satisfies $P_B \leq 10^{-6}$ and the key length is 128 bits for all code points.}
	\label{fig:codecomparisons}
\end{figure*}

%%%%%%%%%%%%%%%%%%%%%%%%%%%%%%%%%%%%%%%%
\subsection{Code Comparisons}
Figure~\ref{fig:codecomparisons} depicts the storage-key $(R_\text{w},R_\text{s})$ projection of the boundary points of the region $\mathcal{R}_{\text{gs},\text{bin}}$ for $p_A=0.15$. Furthermore, we show the point with the maximum secret-key rate $R_\text{s}^*$ and the minimum storage rate $R_\text{w}^*$ to achieve $R_\text{s}^*$. For the FCS and COFE, use the random coding union bound \cite[Thm. 16]{Polyanskiy} to confirm that the plotted rate pairs are achievable for a secret-key length of $128$ bits, an error probability of $P_B=10^{-6}$, and blocklengths of $n=1024$ and $n=2048$. These rate pairs are shown in Figure~\ref{fig:codecomparisons} to the right of the dashed line representing $R_\text{w}+R_\text{s}=1$ bit/source-bit. Similarly, the rate pairs achieved by the previous polar code design in \cite{IgnaPolar}, and Codes 1 and 2 are shown in Figure~\ref{fig:codecomparisons}. 

Storage rates of the FCS and COFE are 1 bit/source-bit, which is suboptimal. The previous polar code construction in \cite{IgnaPolar} achieves a rate point with $R_\text{s}+R_\text{w} =1$ bit/source-bit, which is expected since this is a SW-coding construction. The previous polar code construction improves on the rate pairs achieved by the FCS and COFE in terms of the key vs. storage ratio. Nested polar codes achieve the key-leakage-storage rates of approximately $(0.125, 0.666, 0.666)$ bits/source-bit by Code 1 and $(0.063, 0.315, 0.315)$ bits/source-bit by Code 2, projections of which are depicted in Figure~\ref{fig:codecomparisons}. These rates are significantly better than all previous code constructions for the same parameters and without any private key assumption. Designed nested polar codes increase the ratio $R_\text{s}/R_\text{w}$ from approximately $0.188$ for Code 1 to $0.199$ for Code 2, suggesting to increase the blocklength to obtain better ratios. Code 2 achieves privacy-leakage and storage rates that cannot be achieved by previous methods without applying the method of \textit{time sharing}, since Code 2 achieves privacy-leakage and storage rates of $0.315$ bits/source-bit that are significantly less than the minimum privacy-leakage and storage rates $R^*_{\text{w}}=R^*_\ell=H_b(p_A)\approx 0.610$ bits/source-bit that can be asymptotically achieved by previous methods.

Apply the sphere packing bound \cite[Eq. (5.8.19)]{gallagerbook} to upper bound the key vs. storage rate ratio that can be achieved by SW-coding constructions for the maximum secret-key rate point. Consider $p_A=0.15$, $n=1024$, and $P_B=10^{-6}$, for which the sphere packing bound requires that the rate of the code $\mathcal{C}$ satisfies $R_\mathcal{C} \leq 0.273$. Assume that the key rate is given by its maximal value $R_\text{s} = R_\mathcal{C}$ and the storage rate is given by its minimal value $R_\text{w} = 1 - R_\mathcal{C}$, then we arrive at $R_\text{s}/R_\text{w} \le 0.375$. A similar calculation for $n=2048$ yields $R_\text{s}/R_\text{w} \le 0.437$. These results indicate that there are still gaps between the maximum key vs. storage rate ratios achieved by WZ-coding constructions and the ratios achieved by Codes 1 and 2. The gaps can be reduced by using different nested polar codes that improve the minimum-distance properties, as in \cite{ourPeterSubcode}, or by using nested algebraic codes for which design methods are available in the literature, as in \cite{bizimThomas}. We remark again that such optimality-seeking approaches, e.g., based on information-theoretic security, provide the right insights into the best solutions for the digital era's security and privacy problems. 

%%%%%%%%%%%%%%%%%%%%%%%%%%%%%%%%%%%%%%%%%%%%%%%%%%%%%%%%%%%%%%%%%%%%%%%%%%%%%%%%%%%%%%%%%%%%%%%%%%%%%%%%%%%%%%%%%%%%%%%%%%
\section{Discussions and Open Problems}\label{sec:DiscussionsandOpenProblems}
\begin{itemize}
	\item We want to use low-complexity scalar quantizers after transformation without extra secrecy leakage; however, the decorrelation efficiency metric does not fully represent the dependency between transform coefficients. What is the right metric to use for choosing the transform used in combination with scalar quantizers? Is mutual information between transform coefficients an appropriate metric for this purpose? The choice of the transform should also depend on a reliability metric such as SNR-packing efficiency so that the transform, quantizers, and the error-correction codes can be designed jointly. What is the right reliability metric for this purpose?
	\item It is shown in \cite{bizimtemperature} that the ambient temperature and supply voltage affect the RO outputs deterministically rather than adding extra random noise, which was assumed in the RO PUF literature. What are the right output models for common PUF types, i.e., what are the deterministic and random components, and how are they related?
	\item SRAM PUFs are already used in products. In the literature there is no extensive analysis of the output correlations between different SRAMs in the same device possibly because SRAM outputs are binary and it is difficult to model the correlation between binary symbols. However, SRAM outputs are modeled in \cite{SoftHelper} as binary-quantized sums of independent Gaussian random variables. Is it possible to determine or approximate the correlations between the Gaussian random variables of different SRAMs? If yes, this might be useful for an attacker to obtain information about the secret sequence generated from the SRAM PUF output, which causes extra secrecy leakage.
	\item The  transform-coding approach discussed above results in reliability guarantees for the random-output RO arrays, which considers an average over all ROs manufactured. The worst case scenario is when the transform coefficient value is on the quantization boundary, for which the secret-key capacity is $0$ bit. If one replaces the average reliability metric used above by a lower bound on the reliability of each RO, i.e., a worst-case scenario metric, how would this change the rate of the error-correction code used? For a fixed code, what should be the optimal bound on the reliability of each RO to maximize the yield, i.e., the percentage of ROs among all manufactured ROs for which the worst-case reliability guarantee is satisfied?
	\item Are the WZ problem and the GS model \textit{operationally equivalent} (cf. operational duality)?
	\item Linear block-code constructions discussed above are for uniformly-distributed PUF outputs. Can one construct other (random) linear block codes that are asymptotically optimal for non-uniform PUF outputs? Is it necessary to use an extensions of the COFE for this purpose?
	\item Consider the nested polar code design procedure given above. It is not possible to construct a code with this procedure for $n\leq 512$ since $q*p_A$ is an increasing function of $q$ for any $q\in [0, 0.5]$. Is a nested polar code construction possible  for $n=512$ if one improves the code design procedure and the decoder?
\end{itemize}

%%%%%%%%%%%%%%%%%%%%%%%%%%%%%%%%%%%%%%%%%%

% % \IEEEtriggeratref{6}
\bibliographystyle{IEEEtran}
\bibliography{IEEEabrv,dissrefs}

\end{document}